\documentclass[twocolumn]{aastex63} 

\usepackage{gensymb}
\usepackage{amsmath}
\usepackage{verbatim}
\usepackage{graphicx}

\newcommand{\PSUAA}{Department of Astronomy \& Astrophysics, 525 Davey Laboratory, The Pennsylvania State University, University Park, PA, 16802, USA}
\newcommand{\PSUCEHW}{Center for Exoplanets and Habitable Worlds, 525 Davey Laboratory, The Pennsylvania State University, University Park, PA, 16802, USA}
\newcommand{\PSETI}{Penn State Extraterrestrial Intelligence Center, 525 Davey Laboratory, The Pennsylvania State University, University Park, PA, 16802, USA}
\newcommand{\NOAO}{NSF's National Optical-Infrared Astronomy Research Laboratory, 950 N.\ Cherry Ave., Tucson, AZ 85719, USA}

\received{December 10, 2021}
\revised{February 4, 2022}
\accepted{February 15, 2022}

\submitjournal{AJ}

\shorttitle{NEID Solar Feed}
\shortauthors{Lin et al.}

\graphicspath{{./}{figures/}}

\begin{document}

\title{Observing the Sun as a star: Design and early results from the NEID solar feed}

\correspondingauthor{Andrea S.J.\ Lin}
\email{asjlin@psu.edu}

\author[0000-0002-9082-6337]{Andrea S.J.\ Lin}
\affil{\PSUAA}
\affil{\PSUCEHW}

\author[0000-0002-0048-2586]{Andrew Monson}
\affil{\PSUAA}

\author[0000-0001-9596-7983]{Suvrath Mahadevan}
\affil{\PSUAA}
\affil{\PSUCEHW}

\author[0000-0001-8720-5612]{Joe P.\ Ninan}
\affil{\PSUAA}
\affil{\PSUCEHW}

\author[0000-0003-1312-9391]{Samuel Halverson}
\affil{Jet Propulsion Laboratory, California Institute of Technology, 4800 Oak Grove Drive, Pasadena, California 91109}

\author{Colin Nitroy}
\affil{\PSUAA}
\affil{\PSUCEHW}

\author[0000-0003-4384-7220]{Chad F.\ Bender}
\affil{Steward Observatory, The University of Arizona, 933 N.\ Cherry Ave., Tucson, AZ 85721, USA}

\author[0000-0002-9632-9382]{Sarah E.\ Logsdon}
\affil{\NOAO}

\author[0000-0001-8401-4300]{Shubham Kanodia}
\affil{\PSUAA}
\affil{\PSUCEHW}
\affil{\PSETI}

\author[0000-0002-4788-8858]{Ryan C. Terrien}
\affil{Carleton College, One North College St., Northfield, MN 55057, USA}

\author[0000-0001-8127-5775]{Arpita Roy}
\affil{Space Telescope Science Institute, 3700 San Martin Dr., Baltimore, MD 21218, USA}
\affil{Department of Physics and Astronomy, Johns Hopkins University, 3400 N.\ Charles St., Baltimore, MD 21218, USA}

\author[0000-0002-4927-9925]{Jacob K.\ Luhn}
\affil{Department of Physics \& Astronomy, The University of California, Irvine, Irvine, CA 92697, USA}

\author[0000-0002-5463-9980]{Arvind F.\ Gupta}
\affil{\PSUAA}
\affil{\PSUCEHW}

\author[0000-0001-6545-639X]{Eric B.\ Ford}
\affil{\PSUAA}
\affil{\PSUCEHW}
\affil{Institute for Computational \& Data Sciences, The Pennsylvania State University, University Park, PA, 16802, USA}
\affil{Center for Astrostatistics, 525 Davey Laboratory, The Pennsylvania State University, University Park, PA, 16802, USA}

\author[0000-0002-1664-3102]{Fred Hearty}
\affil{\PSUAA}
\affil{\PSUCEHW}

\author[0000-0003-2451-5482]{Russ R.\ Laher}
\affiliation{IPAC, California Institute of Technology, 1200 E. California Blvd, Pasadena, CA 91125, USA}

\author{Emily Hunting}
\affil{\NOAO}

\author{William R.\ McBride}
\affil{\NOAO}

\author[0000-0002-0289-3135]{Noah Isaac Salazar Rivera}
\affil{Steward Observatory, The University of Arizona, 933 N.\ Cherry Ave., Tucson, AZ 85721, USA}

\author[0000-0002-2488-7123]{Jayadev Rajagopal}
\affil{\NOAO}

\author[0000-0002-7827-6184]{Marsha J.\ Wolf}
\affil{Department of Astronomy, University of Wisconsin-Madison, 475 N.\ Charter St., Madison, WI 53706, USA}

\author[0000-0003-0149-9678]{Paul Robertson}
\affil{Department of Physics \& Astronomy, The University of California, Irvine, Irvine, CA 92697, USA}

\author[0000-0001-6160-5888]{Jason T.\ Wright}
\affil{\PSUAA}
\affil{\PSUCEHW}
\affil{\PSETI}

\author[0000-0002-6096-1749]{Cullen H.\ Blake}
\affil{Department of Physics and Astronomy, University of Pennsylvania, 209 South 33rd Street, Philadelphia, PA 19104}

\author[0000-0003-4835-0619]{Caleb I. Ca\~nas}
\altaffiliation{NASA Earth and Space Science Fellow}
\affil{\PSUAA}
\affil{\PSUCEHW}

\author[0000-0003-0790-7492]{Emily Lubar}
\affil{McDonald Observatory and Department of Astronomy, The University of Texas at Austin, 2515 Speedway, Austin, TX 78712, USA}

\author[0000-0003-0241-8956]{Michael W.\ McElwain}
\affil{Exoplanets and Stellar Astrophysics Laboratory, NASA Goddard Space Flight Center, Greenbelt, MD 20771, USA}

\author[0000-0002-4289-7958]{Lawrence W.\ Ramsey}
\affil{\PSUAA}
\affil{\PSUCEHW}

\author[0000-0002-4046-987X]{Christian Schwab}
\affil{Department of Physics and Astronomy, Macquarie University, Balaclava Road, North Ryde, NSW 2109, Australia}

\author[0000-0001-7409-5688]{Gudmundur Stefansson}
\altaffiliation{Henry Norris Russell Fellow}
\affil{Department of Astrophysical Sciences, Princeton University, 4 Ivy Lane, Princeton, NJ 08540, USA}

\begin{abstract}

Efforts with extreme-precision radial velocity (EPRV) instruments to detect small-amplitude planets are largely limited, on many timescales, by the effects of stellar variability and instrumental systematics. One avenue for investigating these effects is the use of small solar telescopes which direct disk-integrated sunlight to these EPRV instruments, observing the Sun at high cadence over months or years. We have designed and built a solar feed system to carry out ``Sun-as-a-star'' observations with NEID, a very high precision Doppler spectrometer recently commissioned at the WIYN 3.5m Telescope at Kitt Peak National Observatory. The NEID solar feed has been taking observations nearly every day since December 2020; data is publicly available at the NASA Exoplanet Science Institute (NExScI) NEID Solar Archive: \url{https://neid.ipac.caltech.edu/search_solar.php}. In this paper, we present the design of the NEID solar feed and explanations behind our design intent. We also present early radial velocity (RV) results which demonstrate NEID's RV stability on the Sun over 4 months of commissioning: 0.66~m/s RMS under good sky conditions and improving to 0.41~m/s RMS under best conditions.
\end{abstract}


\section{Introduction}
\label{sec:intro}

Since the discovery of the first exoplanets in the early 1990s \citep{Wolszczan_1992, Wolszczan_1994, Mayor_1995}, a remarkable number of planets have been detected, with an astonishing diversity of properties. 
With the Astro2020 Decadal Survey \citep{astro2020decadal} endorsing a flagship ultraviolet/optical/infrared space mission geared toward direct imaging of exoplanets and biosignature detection in exoplanetary spectra, it is imperative to discover and characterize planet candidates in order to increase future scientific yields---especially the ``temperate terrestrial planets orbiting Sun-like stars'' that will be the ultimate prizes of such a mission.
However, an Earth-mass planet orbiting in the Habitable Zone of a solar-mass star only imparts an RV semi-amplitude on the order of 10~cm/s, requiring unprecedented levels of precision and stability which are the goals of a new generation of RV spectrographs. 

NEID \citep{NEID_optical} is among the instruments forging a path towards such extreme RV precision, with a bottom-up error budget yielding an estimate of 27~cm/s single-measurement instrumental precision \citep{NEID_budget}.
In NEID, starlight from the WIYN 3.5m Telescope\footnote{The WIYN Observatory is a joint facility of the NSF's National Optical-Infrared Astronomy Research Laboratory, Indiana University, the University of Wisconsin-Madison, Pennsylvania State University, the University of Missouri, the University of California-Irvine, and Purdue University.} is coupled via a port adapter \citep{NEID_Port_Design, NEID_Port_Overview} into an ultra-stabilized, high-resolution (R $\sim$120,000) fiber-fed spectrograph. NEID covers a broad wavelength range from 380 to 930~nm, and is wavelength-calibrated by an astro-comb (a purpose-built laser frequency comb) and a Fabry-P\'erot etalon.
Other instruments designed to achieve similar precision goals include ESPRESSO \citep{ESPRESSO_main}, EXPRES \citep{EXPRES_main}, MAROON-X \citep{maroonx_main}, and KPF \citep{KPF_main}, while future RV instruments such as HARPS3 \citep{HARPS3_main} and G-CLEF \citep{GCLEF_main} will seek to improve detection sensitivity even further.

Despite such advances in instrumentation, it will be nearly impossible to discover any planets with RV amplitudes of 10~cm/s without first addressing the problem of stellar variability---the EPRV Working Group \citep{Crass2021} recognizes stellar variability as ``the most significant obstacle to achieving EPRV capabilities''.
Stellar processes produce quasi-periodic RV variations (often referred to as ``jitter'') typically on the order of a few m/s \citep{Dumusque_2011}, which would completely mask the RV signal of an Earth-twin exoplanet. Worse still, such RV variations may masquerade as real planets \citep[see for example][]{Santos_2014, Robertson_2014, Lubin2021}, leading to potential false positive detections requiring meticulous vetting. \citet{Fischer_2016} succinctly summarize the current state of the field: ``detailed spectroscopic, photometric, wavelength dependent, and polarization signatures'' offer opportunities for distinguishing 
these stellar effects from the pure velocity shifts induced by planets, but finding and utilizing these signatures is still a critical area of ongoing research. The promise of these techniques is yet to be realized, and extensive data sets of solar RVs from multiple instruments can play a key role.

The Sun, our closest star, is an ideal test case for studying RV variations from both stellar and instrumental sources, because it is the only star whose surface can be resolved in great detail. The Sun is particularly advantageous because we can compare disk-integrated RVs with disk-resolved observations from other solar instruments \citep{Haywood_2016, Thompson_2020}, allowing us to test our models and predictions. In addition, since solar light usually follows a similar light path through the instrument as starlight, it serves as a valuable diagnostic tool for revealing and tracking instrumental systematics \citep{CollierCameron_2020, Dumusque_2020}, especially when leveraging simultaneous solar observations from multiple RV facilities.

Stellar astrophysical processes that can affect RVs include, but are not limited to: convection-driven phenomena (p-mode pulsation, granulation, and supergranulation), starspots, plage, flares, coronal mass ejections, and long-term magnetic activity cycles.
Observing strategies for suppressing RV noise from stellar p-modes---acoustic standing waves which cause the star to ``ring'' on timescales of many minutes---are relatively well-understood \citep{Chaplin_2019}, although they have not been validated with real observations at precisions of a few cm/s. Strategies for reducing the RV effects of granulation have also been validated \citep{Dumusque_2011, CollierCameron_2019}, though compared to p-mode mitigation strategies they are less effective at lowering the overall RV RMS. 
However, mitigation of supergranulation is still a challenge \citep[see for example][]{Meunier_2019}, in part because the phenomenon is not completely understood.

For magnetic activity in quiet stars, the largest source of RV variability is usually the suppression of convective blueshift by magnetically active regions, which distorts the disk-integrated line shapes.
For solar spectral lines, we know that such effects vary as functions of wavelength and line depth, with some lines being more magnetically sensitive than others \citep{Meunier_2017b}. This serves as the foundation of many different approaches to addressing activity-induced RVs. Activity-correlated lines can be identified by looking for lines whose measured RVs do not change in the same way as the general line population, so that they can be dropped from the set used to derive precise RVs and/or leveraged for more detailed analysis \citep{Dumusque_2018, Rajpaul_2020, Cretignier_2020}. Meanwhile, other efforts are centered around the development of better activity indicators, whether these be magnetically sensitive lines \citep{Maldonado_2019, Thompson_2020} or other stellar properties like the unsigned magnetic flux \citep{Haywood_2020}. However, further work is still required in this area---with an emphasis on the need for distinct indicators for different types of magnetic activity, such as plage versus chromospheric network \citep{Milbourne_2019}---as merely addressing suppression of convective blueshift on a global scale is insufficient for noise mitigation on magnetically quiet stars where it may not dominate contributions to RV activity \citep{Miklos_2020}.

Yet other methods focus on tracking stellar-induced RV variability through time series. Gaussian process models can be used to jointly model RVs in conjunction with existing activity indicators to find planets despite the confounding effects of stellar activity \citep{Rajpaul_2015, Jones_2017}, for example, the CoRoT-7 system investigated by \citet{Haywood_2014}.
The analysis of three years of \mbox{HARPS-N} solar data has shown that the FWHM of the cross-correlation function as well as its asymmetry, as quantified by the bisector inverse slope, can be used to track solar RVs---albeit with complicated time lags \citep{CollierCameron_2019}.
With sufficient quantities of data, it is possible to explore methods for disentangling velocity shifts from line-shape changes, for example, by examining the principal components of the auto-correlation function \citep{CollierCameron_2020}, or by using artificial neural networks to identify line-shape changes without relying on time-domain information \citep{deBeurs_2020}.

The usefulness of disk-integrated Sun-as-a-star data has already been proven by the solar telescope at \mbox{HARPS-N} \citep{Dumusque_2015, Phillips_2016}, which has produced a rich and homogeneous data set over five years of operation.
We expect that solar data with NEID will be especially valuable due to its extended red-optical/near-infrared (NIR) coverage out to 930~nm (while many other RV spectrographs stop around 700~nm). The impact of stellar processes on RVs can be significantly lower in the NIR, as the difference in relative flux contributions from magnetically active versus magnetically quiet regions is reduced at these longer wavelengths.
Furthermore, the NIR offers possibilities for directly measuring stellar magnetic fields via Zeeman splitting, since the effect is more pronounced at longer wavelengths, and features such as the calcium NIR triplet (8498, 8542, and 8662 \AA) can serve as robust activity indicators \citep{Marchwinski_2015, Robertson_2016}.

In this paper, we describe the design, integration, and operations of the NEID solar feed, along with solar RV measurements from NEID instrument commissioning. In Section \ref{sec:design}, we outline our design choices for the NEID solar feed and the design intent behind these choices. Section \ref{sec:integration} describes the integration of the NEID solar feed at WIYN. In Section \ref{sec:observations}, we discuss our observing strategy, data processing, and current database of solar data. Finally, we present RV observations of the Sun obtained during NEID commissioning in Section \ref{sec:early_results}.

\section{Design of the NEID Solar Feed}
\label{sec:design}

The timeline for the development and construction of the NEID solar feed was aggressive, as the solar feed was a late addition to the instrument, and we wanted it available in time to support the final stages of NEID spectrograph testing at Pennsylvania State University (Penn State). Therefore, we sought to utilize readily-available, commercial off-the-shelf (COTS) components wherever possible, which allowed us to design and build the full solar feed system in less than 6 months.

We loosely based the initial design of the NEID solar feed on the \mbox{HARPS-N} and GIARPS solar telescopes (\citet{Phillips_2016} and \citet{Claudi_2018}, respectively), as these instruments have demonstrated the ability to survive outdoor conditions for years while producing high-quality solar data. The optics in these telescopes consist of a single small lens which feeds sunlight into an integrating sphere. The integrating sphere provides high levels of spatial scrambling, producing disk-integrated sunlight---mimicking the light received from an exoplanet host star---which is then coupled into an optical fiber that feeds the spectrograph.

\begin{figure}[!htbp]
    \centering
    \includegraphics[width=0.45\textwidth]{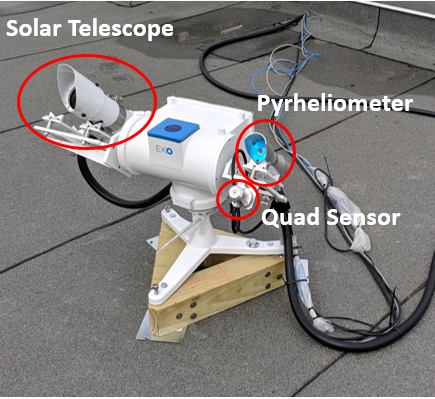}
    \caption{The solar feed assembly, shown during NEID spectrograph testing at Pennsylvania State University (Penn State). One leg of the solar tracker (pointing rightwards in this image) is approximately aligned with geographic north, and adjustable screw-feet in each leg are used to level the tracker. The wooden base was a temporary addition for stability, as we could not bolt the tracker down at this time. Our custom-built optics assembly is affixed to the shelf mount on the left side of the tracker, with the optical fiber leading to the spectrograph protected by the thick black conduit. On the other arm of the tracker are the pyrheliometer and the quadrant sensor which drives the tracker's active guiding loop. The optics assembly and the pyrheliometer are co-aligned with the quadrant sensor by using pinhole sights, adjusted by screws in the shelf mounts.}
    \label{fig:solar_tel_labeled}
\end{figure}

\begin{figure*}[!htbp]
    \centering
    \includegraphics[width=0.8\textwidth]{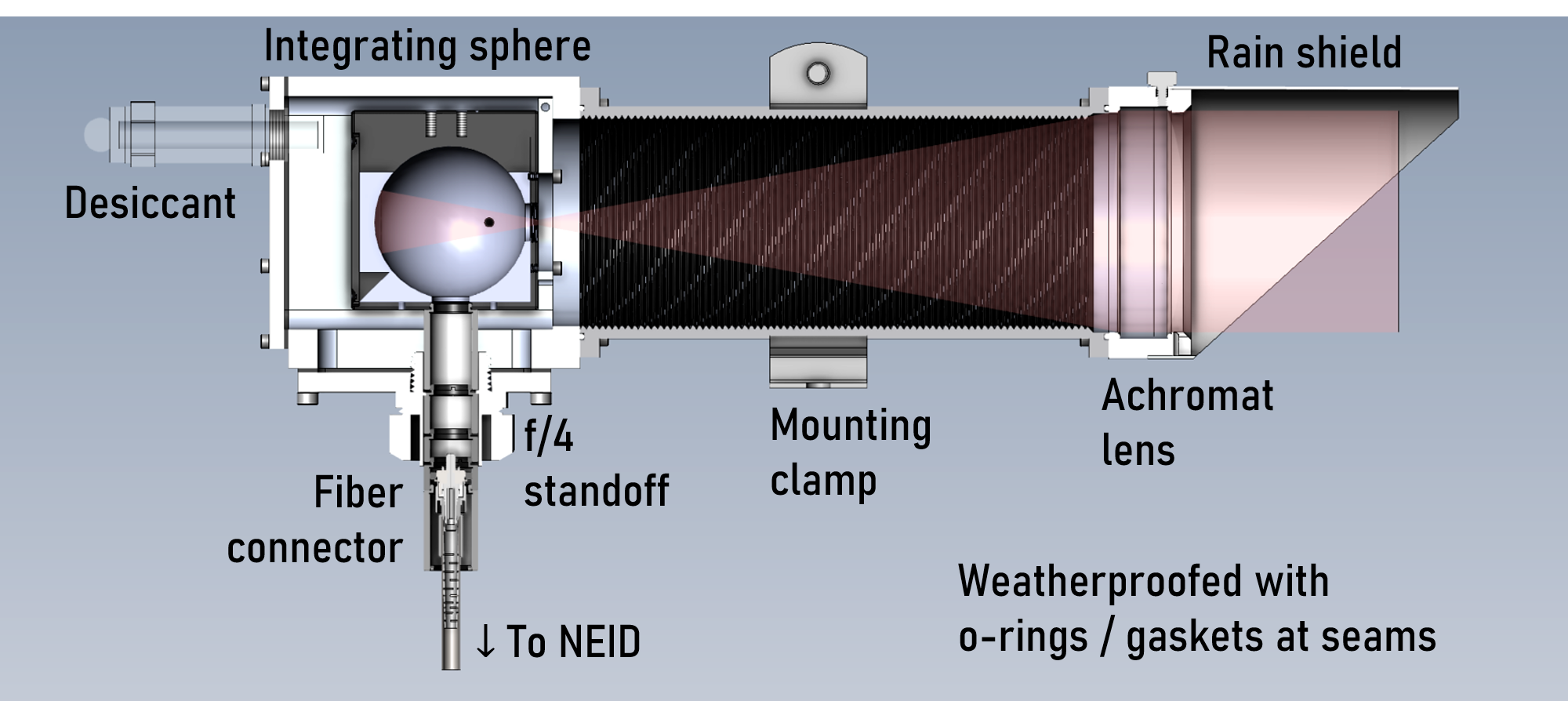}
    \caption{A cutaway view of the solar telescope optics assembly, rendered in SolidWorks. The pink shading represents the approximate beam of the incoming sunlight. All parts except for the outer housing were commercial off-the-shelf (COTS) components.}
    \label{fig:telescope_cutaway}
\end{figure*}

\autoref{fig:solar_tel_labeled} shows the NEID solar feed assembly. A COTS solar tracker maintains stable pointing, via an active feedback loop driven by a quadrant sensor. The optics assembly, shown in detail in \autoref{fig:telescope_cutaway}, contains the lens and integrating sphere. We also use a pyrheliometer as a dedicated cloud sensor. This is an instrument commonly used in photovoltaic science to monitor solar flux (specifically direct normal irradiance, DNI) over a large wavelength range, but for our purposes, it fulfills the role of directional cloud sensor, since obscuration of the solar disk will cause a drop in flux. 

The spatially-scrambled sunlight travels through $\sim$45 meters of optical fiber to a dedicated shutter mechanism on the NEID calibration bench. The light is then directed to the spectrograph science fiber via the NEID port adapter, using the path for internal fiber-to-fiber calibration (see Subsection \ref{sec:path} for more details). This permits the use of a simultaneous calibration source, through the normal calibration fiber, for maximum RV precision.

A full list of the major components of the NEID solar feed and their costs can be found in \autoref{tab:parts_table} in the Appendix.

{}

\subsection{Lens}
\label{sec:lens}

In choosing a lens for our solar telescope, our primary concern was to collect enough light---even with the Sun as our source!---because the solar light must pass through an integrating sphere and several long fiber runs before reaching the rest of the instrument (see Subsection \ref{sec:integrating_sphere} for more details). An excess of light can be addressed with a neutral-density filter or an aperture stop, but a deficit is much harder to solve. 

An important secondary consideration was to ensure that the incoming sunlight passed through the input port of the integrating sphere at all wavelengths of interest. To help with this, we chose to place the focal point of the lens at the center of the open input port. This yields the smallest possible beam size, with the focused solar image 1.8~mm in diameter, thus minimizing the chance of truncating the solar image due to tracking error. However, the collected sunlight is highly concentrated, so it is imperative that the focal point is located in free space.

We considered several COTS lenses (summarized in \autoref{tab:lens_table}), but ultimately chose one very similar to the \mbox{HARPS-N} design---a 75~mm diameter achromatic doublet with a 200~mm focal length (Edmund Optics 88-596). We chose the Edmund Optics proprietary \mbox{VIS-NIR} anti-reflection (AR) coating instead of MgF$_2$, providing us with lower reflectance across a larger fraction of NEID's wavelength range, though at the cost of more structure in the reflectance curve. 

\begin{table*}[!htbp]
\centering
\begin{tabular}{lllllll}
\hline
Manufacturer/Part \# & Dia. (mm) & EFL (mm) & Material     & AR Coating & T$_{380}$ & T$_{635}$ \\ \hline
\textbf{Edmund Optics 88-596} & \textbf{75}            & \textbf{200}      & \textbf{N-BK7/N-SF5}  & \textbf{VIS-NIR}    & \textbf{0.753}      & \textbf{0.987}      \\
Edmund Optics 45-417 & 75            & 200      & N-BK7/N-SF5  & MgF$_2$       & 0.718      & 0.942      \\
Edmund Optics 33-925 & 75            & 150      & N-BAK1/N-SF8 & VIS-NIR    & 0.391      & 0.974      \\
Thorlabs AL50100G    & 50            & 100      & N-BK7        & None       & 0.915      & 0.919      \\ \hline
\end{tabular}
\caption{Summary of properties for lenses considered for the NEID solar feed. $T_{380}$ and $T_{635}$ refer to the total transmittance (substrates + AR coating) at the specified wavelengths (in nm). We chose the first lens due to its larger light-collecting area, consistent spot size, and better throughput---both overall and at the bluest wavelengths where instrumental throughput is already low.}
\label{tab:lens_table}
\end{table*}

\begin{figure*}[!htbp]
    \centering
    \includegraphics[width=\textwidth]{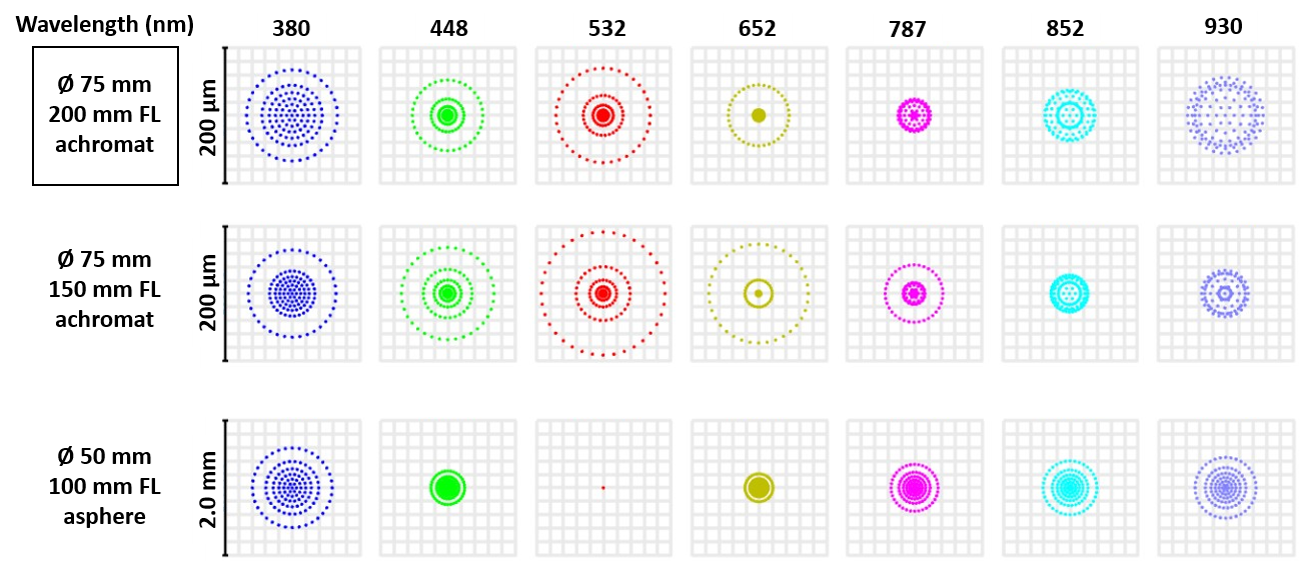}
    \caption{Zemax OpticStudio comparison of focus spot sizes of several candidate lenses over the NEID wavelength range (given across the top in nm). We chose the lens depicted in the top row, which has relatively consistent spot sizes across our wavelengths of interest. Note the 10x change in scale for the asphere (bottom row). We were concerned that the large size of these spots at the extremes of our wavelength range greatly increased the chance that the edges of the solar disk might fall outside the integrating sphere input port.}
    \label{fig:lens_comparison}
\end{figure*}

We initially considered using an aspheric lens in order to minimize the effects of spherical aberration, but decided that the drawbacks outweighed the benefits. \autoref{fig:lens_comparison} shows that the focus spots produced by an aspheric lens grow quickly in size away from the design wavelength. Given NEID's broad wavelength coverage and our desire to keep the spot size relatively uniform across the whole NEID bandpass, this was not ideal. In addition, the largest COTS asphere we found that met our tolerances for surface smoothness was only 50~mm in diameter (the Thorlabs AL50100G in \autoref{tab:lens_table}), and did not have a readily-available AR coating with acceptable transmission across our entire wavelength range.

We also desired a shorter focal length, to decrease the overall size and weight of the optics assembly, but found that the glass used to achieve these shorter focal lengths sharply drops in transmission efficiency at the bluemost extent of the NEID wavelength range ($<$~400~nm). We were unwilling to accept further losses where the instrumental throughput of NEID is already lowest, as this wavelength region contains important solar activity information in the form of the Ca~II H \& K lines.

Since the solar feed assembly will remain outdoors for years, we wanted to ensure the optics would survive intense solar light and a variety of weather conditions. We are not concerned about the doublet nature of our chosen lens, because the differential thermal expansion of the two substrates is negligible under normal environmental temperatures. Furthermore, the epoxy used to bond them together is Norland Optical Adhesive 61 (NOA-61), which can withstand temperatures from $-150\degree$ to $+125\degree$C and does not solarize with exposure to ultraviolet light (UV tolerant). Similarly, the VIS-NIR coating is non-hygroscopic, abrasion-resistant, unaffected by solar UV, and can be cleaned with isopropyl alcohol. Furthermore, if a replacement lens is necessary, it is a COTS component which can be easily acquired. Since the solar feed assembly was installed at WIYN in November 2019, no noticeable damage or degradation of the lens has occurred.

\subsection{Integrating Sphere}
\label{sec:integrating_sphere}

Our primary concern when choosing an integrating sphere was the size of the sphere itself---too large a sphere might fail to provide enough output flux, while too small a sphere might not provide sufficiently uniform scrambling (due to a larger fraction of its area being taken up by input/output ports). Thus, we followed the \mbox{HARPS-N} design, which has been proven to work well. This 2-inch PTFE sphere, manufactured by Thorlabs, provides uniformly high reflectivity ($\sim$99\%) throughout the entire NEID wavelength range and is compatible with many other optics parts, making it easier to make modifications or attach additional components.

We verified that the power density of the incoming sunlight would not damage the interior of the integrating sphere during normal operation. With a 75~mm diameter lens and a 200~mm focal length, a geometric estimate yields roughly 2.2~W/cm$^{2}$ on the integrating sphere surface opposite the input port, assuming the airmass-zero solar constant of $\sim$1361~W/m$^{2}$. Even before accounting for atmospheric losses, this is \emph{much} lower than the maximum sustained power density of 2000~W/cm$^{2}$ for which the sphere is rated.

We also checked that a small tracking error would not displace the focal point outside the input port, potentially damaging the sphere housing. We confirmed with a Zemax OpticStudio analysis of our telescope optics that even with a displacement of 2 full solar diameters ($\sim$1$\degree$) off-center, the solar image still stays solidly within the bounds of the integrating sphere port.
Combined with the tracking accuracy of our system (discussed below in Subsection \ref{sec:solar_tracking}), we are confident that we do not need to worry about damage to the sphere housing.

\subsection{Fiber Feed}
\label{sec:fiber_feed}

The solar fiber is captured in a standard FC/PC connectorized end, set back from the output port of the integrating sphere by a standoff tube, as seen in \autoref{fig:telescope_cutaway}. The length of the standoff tube is adjusted to achieve an input beam of approximately \textit{f}/4, to match the \textit{f}/4 input used by the rest of the instrument.

\begin{figure}[!htbp]
    \centering
    \includegraphics[width=0.45\textwidth]{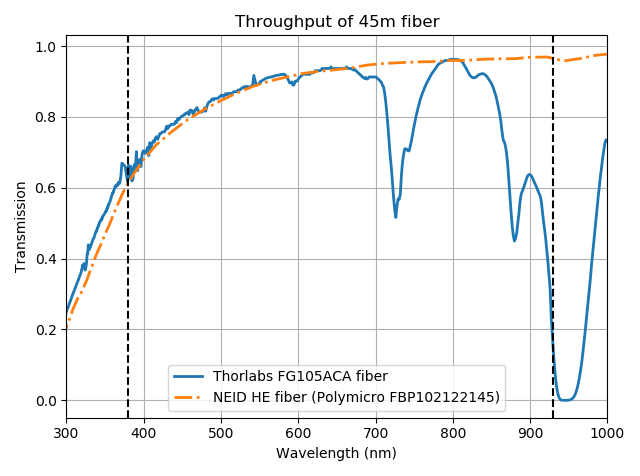}
    \caption{Transmission as a function of wavelength for equivalent lengths (45m) of the Thorlabs FG105ACA fiber used at Penn State and the NEID HE fiber (Polymicro FBP102122145) used at WIYN. The NEID wavelength range of 380~nm to 930~nm is marked with vertical dashed lines. The NEID HE fiber shows much less structure in its transmission curve, as well as better transmission in the red.}
    \label{fig:fiber_thru_comp}
\end{figure}

During initial testing at Penn State, we used a $\sim$40-meter length of Thorlabs FG105ACA solarization-resistant fiber to connect the solar feed assembly (located on the roof above the NEID integration clean room) to the spectrograph. At this time, the FG105ACA fiber was our best available option, despite the prominent structure in the transmission curve shown in \autoref{fig:fiber_thru_comp}, because we were concerned about loss of blue throughput due to solarization. 

For final installation of the solar feed at WIYN, we used the same Polymicro FBP102122145 105-micron multimode fiber used for the NEID High Efficiency (HE) mode, which has a much smoother transmission curve. We procured two custom-jacketed 45-meter lengths of this fiber, assembled by FiberTech Optica. The jacketing consists of stainless steel wrap within a layer of black PVC. At our request, the ends of these fibers were built with plastic connectors instead of the standard metal ones---due to the risk of lightning strike at WIYN, we wanted to avoid a direct electrically conductive path between the exterior solar telescope assembly and any NEID or WIYN electronics. One of these custom fibers has been installed in the solar feed, while the other one is kept in storage at the observatory as backup.

\subsection{Solar Tracking}
\label{sec:solar_tracking}

Due to the accelerated timeline of the NEID solar feed, we did not have the time or personnel to dedicate to creating a custom tracking system, such as the two CMOS cameras that maintain the pointing of the \mbox{HARPS-N} solar telescope \citep{Phillips_2016}. We also investigated amateur astronomy mounts with solar tracking functionalities, but these are typically not designed for long-term autonomous use.

Fortunately, there is already a market for rugged, autonomous solar trackers in photovoltaic and environmental science fields. We acquired an EKO Instruments STR-22G solar tracker, which is designed to operate autonomously even in harsh environments, including remote operations in Antarctica \citep{EKO_antarctica}, and optimized to track the motion of the Sun. We chose this particular tracker because it had the most options for mounting our optics assembly and additional sensors, and because it had both active and passive tracking modes. In the active tracking mode, a quadrant sensor is used to keep the solar tracker centered on the Sun's position. When there is insufficient flux in the quadrant sensor to provide accurate centering, such as when clouds obscure the Sun, the tracker falls back to passive tracking where it uses GPS location to predict the Sun's position based on an internal model.

\subsubsection{Tracking Test}

\begin{figure*}[!htbp]
    \centering
    \includegraphics[width=\textwidth]{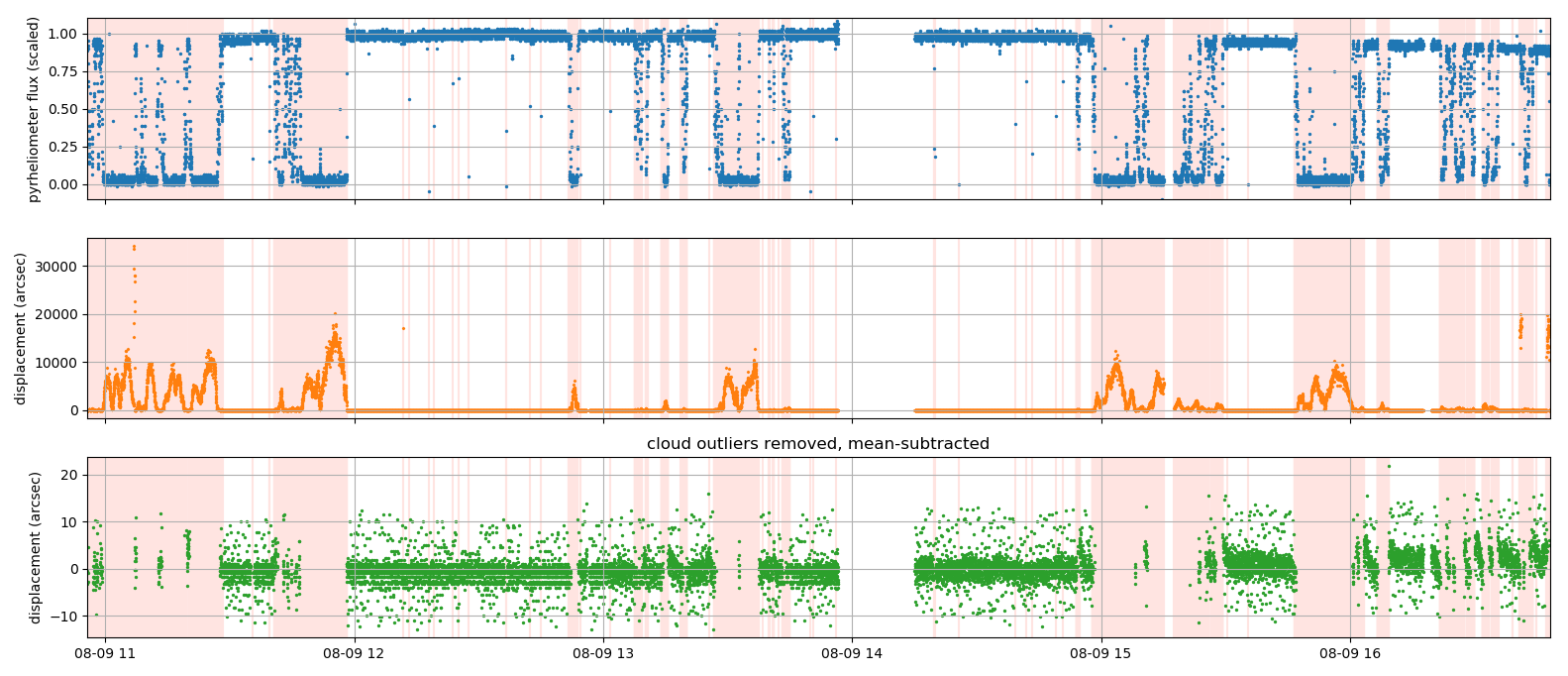}
    \caption{Scaled pyrheliometer flux (\textit{top panel}) and position sensor centroid (\textit{middle}) during the solar tracking test. A short span of no data after 14:00 is due to the repositioning of the solar tracker. Periods where the measured centroid deviates from the center of the sensor by more than a few arcseconds correlate strongly with low flux in the pyrheliometer, which we take to be the result of clouds or other obscuration of the Sun. When these cloudy periods---marked by pink shading---are removed by a flux cut discarding points with values $< 80\%$ of the normalized flux (\textit{bottom}), the tracking is accurate to a RMS of 3.2~arcsec over 6~hours. We believe the deviation from zero displacement seen towards the end of the clear tracking graph is a chromatic effect, due to the quadrant sensor and our imaging system operating at different wavelengths.}
    \label{fig:tracking_test}
\end{figure*}

The STR-22G solar tracker is stated to have a ``tracking accuracy'' of $\sim$0.01$\degree$ (36~arcsec). To verify this, we attached a small imaging system to the tracker to monitor the location of the solar image on a position sensor. This assembly consisted of a lens, a reflective neutral density filter (to reduce the intensity of the sunlight), and a Thorlabs PDP90A 2D lateral position sensor. We set up the tracker according to its instructions and let it operate autonomously for $\sim$6~hours on a partly cloudy day. The results of this test are shown in \autoref{fig:tracking_test}.

Under clear sky conditions, the STR-22G solar tracker actually performed much better than its specifications, maintaining its pointing on the Sun to a RMS of 3.2~arcsec ($\sim$0.2\% of a solar diameter), with typical excursions out to 10~arcsec. During cloudy periods, as indicated by low flux in the pyrheliometer, the centroid measured by the position sensor wandered by a degree or more. However, we believe that because this effect is so strongly correlated with clouds, it \textit{does not} reflect the tracker's true accuracy and is merely a consequence of uneven extinction of the Sun, with the true tracking accuracy better reflected by the 3.2~arcsec RMS observed during clear conditions.

The solar tracker had to be physically picked up and relocated halfway through the tracking test in order to keep it in direct sunlight, which resulted in a short interruption in the data stream. Conveniently, this allowed us to confirm the robustness of the tracking algorithms---we observe no difference in tracking accuracy before and after this interruption, demonstrating that the tracker's accuracy is unaffected by physical shifts or rotations, as long as the Sun remains within the quadrant sensor's field of view.

As demonstrated by our tracking test, the RMS error of our solar tracker is 3.2~arcsec and the largest observed deviations under clear conditions were $\ll$~1~arcminute. These deviations are much smaller than the 6~arcmin accuracy reported by \citet{Phillips_2016} for 10~cm/s RV precision with the \mbox{HARPS-N} solar telescope. Thus, we expect our tracking accuracy to have a negligible contribution to any observed RV variations, although we have not explicitly tested the on-sky RV impact of a misalignment.
In summary, this COTS solution met and exceeded our needs for accuracy and precision in a solar tracker.

\subsection{Flux Monitoring}
\label{sec:flux_monitoring}

As demonstrated by our tracking test, uneven extinction of the solar disk---for example, due to passing clouds---can cause large spurious RV shifts. Because the Sun is a resolved source, heterogeneous obscuration of the solar disk will preferentially block either redshifted or blueshifted light, distorting the observed RVs. 

Under normal operating procedures, the NEID spectrograph relies on its internal chromatic exposure meter, which operates at a 1~Hz cadence, to indicate clouds or poor seeing during observations. However, one of the purposes of the solar feed during spectrograph testing was to serve as an independent check of the exposure meter, so we included a second, more direct, method to detect drops in solar flux.

A standard astronomical all-sky camera did not suit our needs. The vast majority of such cameras are designed to operate at night only and thus do not cope well with the Sun; they also measure full-sky cloud cover, whereas we only care about clouds in a small radius around the Sun. Furthermore, with an image-based cloud sensor, it can be difficult to accurately identify the presence of thin clouds or hazes.

Instead we use a pyrheliometer, an instrument commonly used in photovoltaic science. The MS-57 pyrheliometer from EKO Instruments measures the total solar direct normal irradiance (DNI) from 200 to 4000~nm in a 5$\degree$ field of view around the Sun---for our purposes, it is effectively a directional bolometer. A thermopile within the pyrheliometer reads out an analog voltage to a LabJack data acquisition module, and the voltage can be converted to an intensity via a factory-calibrated conversion factor. We set the readout rate of the pyrheliometer to 1~Hz to match the cadence of the NEID exposure meter, as faster cadences offer no significant benefit for detecting clouds during solar exposures.

\begin{figure*}[!htbp]
    \centering
    \includegraphics[width=0.9\textwidth]{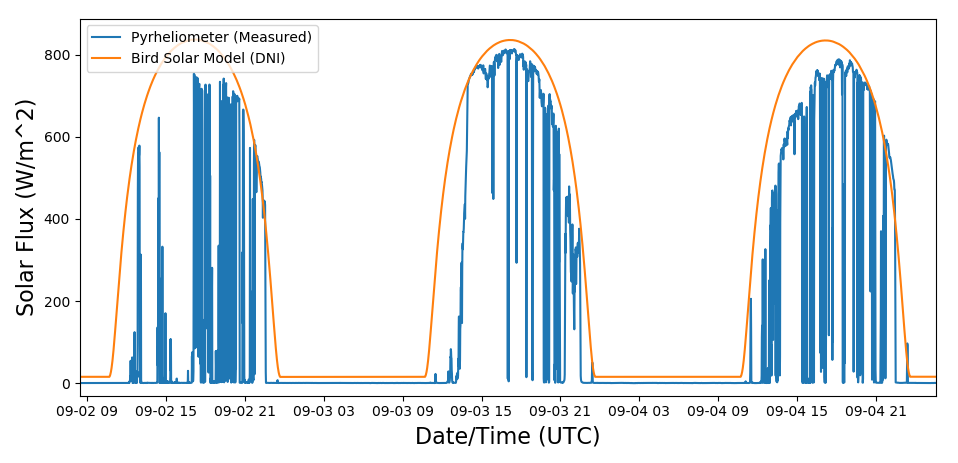}
    \caption{Comparison of measured pyrheliometer flux with the Bird Clear Sky Model over 3 days during NEID testing at Penn State, showing significant clouds. We expect some small discrepancies in predicted vs.\ measured flux during clear periods, as we estimated the local atmospheric water vapor and aerosol levels for the Bird Model.}
    \label{fig:pyr_vs_bird}
\end{figure*}

The DNI over the course of any given day can be calculated through models such as the Bird Clear Sky Model \citep{Bird_1981}; we use the Python implementation found in \texttt{pvlib} \citep{Holmgren2018_pvlib}. Deviations from the predicted flux thus indicate the presence of clouds or other obscuring phenomena; \autoref{fig:pyr_vs_bird} shows 3 days of typical weather at Penn State during NEID spectrograph testing. The DNI from the Bird Model depends on atmospheric properties---such as water vapor and aerosol content---which we must estimate, but even an approximation gives us a quick and effective tool to gauge the overall quality of our solar observations.

\subsection{Weather Enclosure}
\label{sec:weather_enclosure}

It is imperative to protect the solar feed assembly against the elements. The WIYN observatory experiences heavy rain and lightning storms during the summer monsoon season, snow and ice during the winter, and frequent high winds due to its position at the end of the mountain ridge. 

We considered enclosing the entire assembly in an acrylic dome, like the \mbox{HARPS-N} solar telescope, or a box with a remote-controlled lid, like LOCNES and HELIOS (the solar telescopes for GIARPS and HARPS, \citet{Claudi_2018} and \citet{HARPS_HELIOS} respectively). However, we were concerned that imperfections in the dome might cause optical aberrations which would differ over the course of the day, producing spurious RV shifts. Similarly, if the focal point of the optics assembly is not at the center of the dome, RV shifts may be induced by the differing angle and curvature of the dome with respect to the other optics. We concluded that in our situation, a box had the potential to create more problems than it would solve: the lid mechanism would be an additional point of failure on a system which might not be easily accessible, and rain and snow could collect inside without careful drainage. The size of the dome or box would also make it more difficult to find a suitable location to mount the solar feed assembly.

Ultimately, we chose to utilize ruggedized and highly weather-resistant components instead of enclosing the solar feed assembly. We designed a custom housing for the lens and integrating sphere---seen in \autoref{fig:solar_tel_labeled} and \autoref{fig:telescope_cutaway}---which was machined by Hilltop Technology Laboratory. The seams are made weather-tight by sandwiching rubber o-rings in the junctions along the tube assembly. An attached desiccant cartridge keeps the interior components dry, and the outside is powder-coated white to resist corrosion. The inside of the main tube is threaded to reduce ghosting from scattered light and blackened with MH2200 paint. 
We also inserted a small Dracal USB sensor at the side of the main tube to monitor the interior temperature and humidity. 

We chose to have the surface of the achromatic lens exposed to the outside, as the \mbox{VIS-NIR} AR coating is non-hygroscopic and abrasion-resistant, and the lens is relatively inexpensive if a replacement is necessary. A flat protective window in front of the lens would cause fringing, and a COTS wedged window of sufficient size was not available. We added a rain shield to the housing to help keep water and snow off the front face of the lens, which might leave behind residue upon evaporating.

The solar tracker, pyrheliometer, and associated cables are already designed to survive unattended in harsh environments. The solar fiber (already jacketed with PVC and steel wrap) is fed through a length of weatherproof flexible rubber conduit, which is attached to the solar telescope housing and runs all the way through to the interior of the WIYN building so that the fiber jacketing is never exposed to the outside.

The only necessary maintenance for the outdoor components of the solar feed assembly is to inspect and clean the lens every few months with a microfiber cloth and isopropyl alcohol, and change the desiccant cartridge if it has become saturated, as indicated by the desiccant within turning pink.

\section{Integration at the WIYN 3.5m Telescope}
\label{sec:integration}

\subsection{Location}
\label{sec:location}

When selecting a location for the solar feed assembly, our goal was to mount it with an unobstructed view of the sky and out of the potential paths of observatory staff or vehicles, while remaining accessible for periodic maintenance. 

We considered mounting the solar feed assembly on the south wall of the dome, where the fiber run to reach the NEID calibration bench could be shortened by $\sim$20~meters. This would mean less attenuation of the incoming sunlight, especially at our bluest wavelengths ($\sim$380 to 400~nm). However, it became apparent that this was not a feasible location, as it would encroach upon a narrow emergency-egress route, and incur the risk of wintertime damage from ice sheets sliding off the dome.

\begin{figure}[!htbp]
    \centering
    \includegraphics[width=0.45\textwidth]{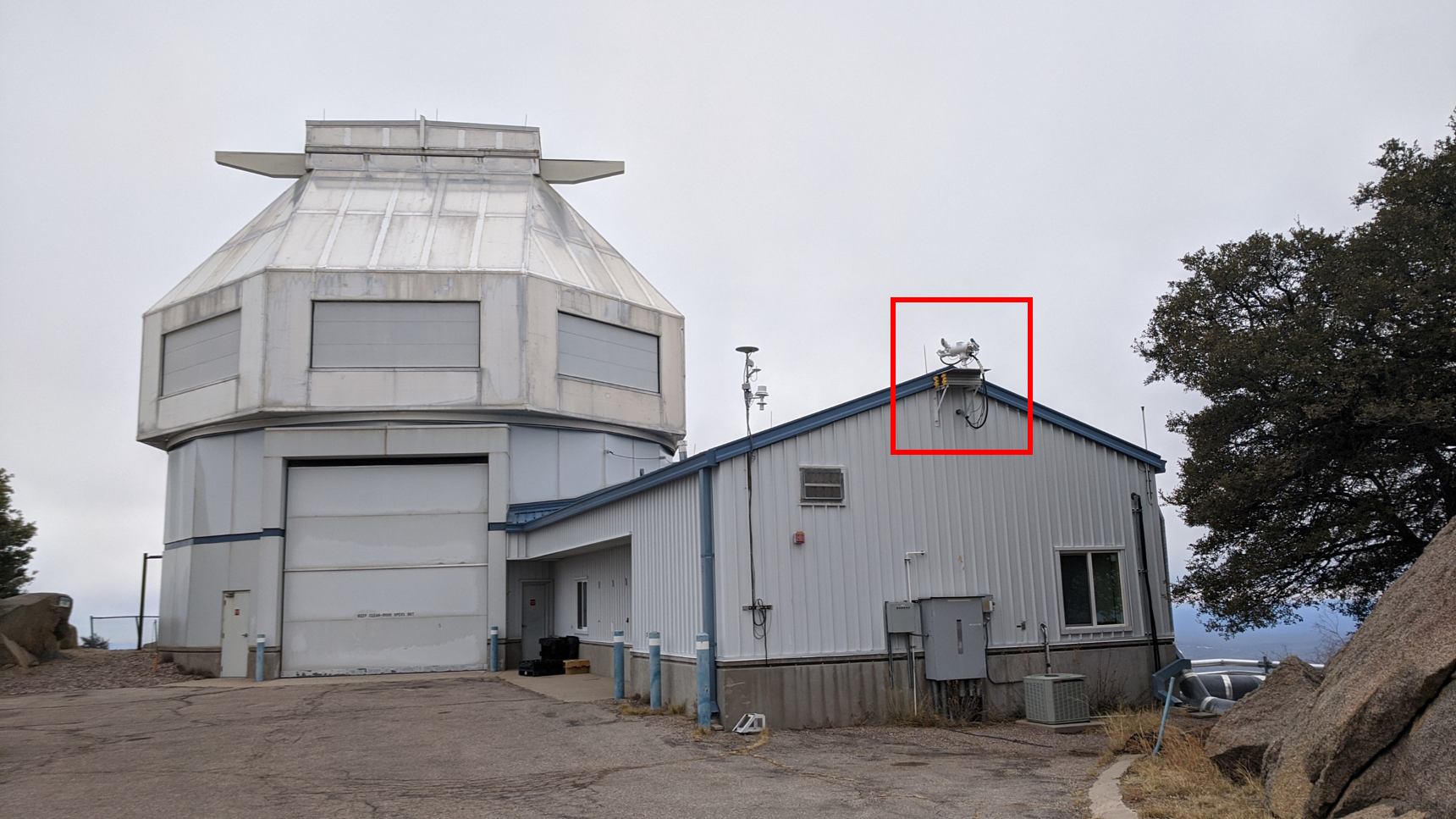}
    \includegraphics[width=0.45\textwidth]{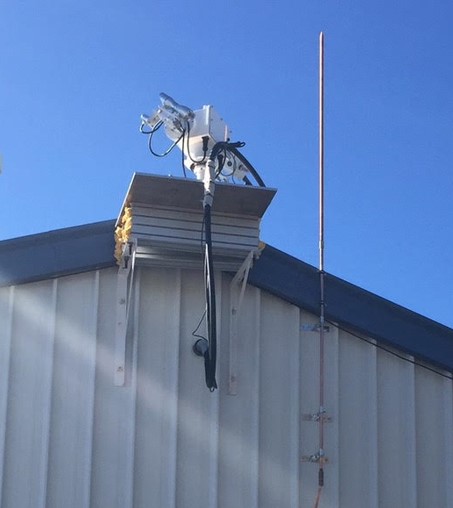}
    \caption{\textit{Top}: The solar feed assembly installed on its shelf on the end of the WIYN control room building. The weather station to its left is short enough that it does not block the solar telescope's view of the sky. A new, taller lightning rod has also been installed near the apex of the roof (after this picture was taken---see bottom panel). \textit{Bottom}: Close-up of the installed solar feed assembly, with the pyrheliometer and quadrant sensor visible on the left tracker arm. Our optics assembly is on the right arm, behind the body of the tracker. The thick black conduit contains the jacketed solar fiber. Cables are affixed to the back of the tracker to reduce the load on the altitude drives, and wrapped in a tough plastic sheeting to prevent wear from rubbing against the edge of the platform as the tracker rotates. The PVC fitting that routes cables inside the building can be seen directly under the shelf, and the new lightning rod is visible to the right of the solar tracker.}
    \label{fig:telescope_installed}
\end{figure}

As a result, the solar feed assembly is mounted on a small shelf attached to the east end of the control room building near the WIYN weather station, as shown in the top panel of \autoref{fig:telescope_installed}. We worked with NOIRLab staff on the design of the shelf, to ensure that it would be robust enough to support the cantilevered weight of the tracker and its accessories even with heavy wind-loading, as well as giving it ample room to rotate over the course of the day. Currently, access to the solar feed assembly for inspection and cleaning is only possible via the mountain's boom lift.

The shelf itself is made of aluminum and is bolted into a building stud near the apex of the roof. A platform was added atop the primary shelf to ensure that the solar telescope sits above the roof level and thus has a nearly-unobstructed view of the sky above airmass 2. The solar tracker is bolted to the top platform via mounting holes in the legs and leveled using adjustable screw-feet. It is about 9$\degree$ misaligned from geographic north, as it was accidentally installed aligned with magnetic north instead. However, the Sun is still within the 30$\degree$ field of view of the sun sensor, so the tracker software automatically compensates for the offset.

When connecting the solar feed assembly, we were careful to leave sufficient slack in the cables to allow for the azimuthal and altitudinal motion of the solar tracker, as seen in the bottom panel of \autoref{fig:telescope_installed}. The solar fiber and associated cables were then routed inside the WIYN control building via an elbow fitting filled with foam to isolate the inside and outside environments.

The location atop the control room has a clear view of the sky above airmass $\sim$2 throughout the year. This yields an average of 6.5 hours of Sun (above airmass 2) per day, with a minimum of $\sim$3 hours in the winter and a maximum of $\sim$9 hours in the summer. However, due to the constraints of our observing strategy (outlined in Subsection \ref{sec:observing_strategy} below), we expect a consistent 6 hours of solar data per day, even in the summer.

\subsection{Weatherproofing}
\label{sec:weatherproofing}

\begin{figure}[!htbp]
    \centering
    \includegraphics[width=0.45\textwidth]{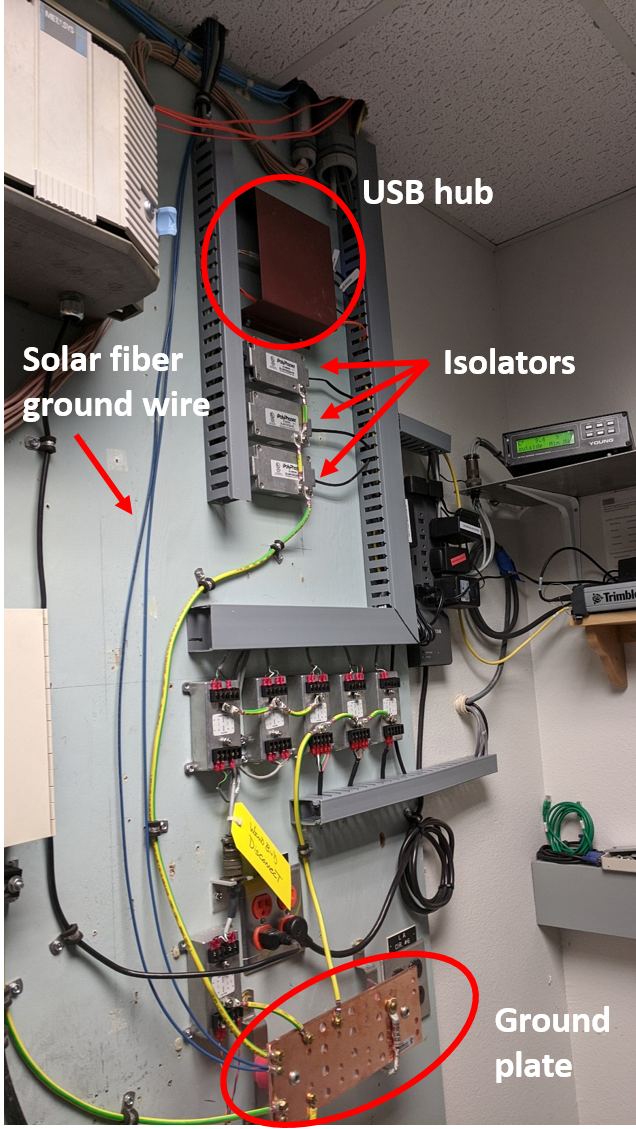}
    \includegraphics[width=0.45\textwidth]{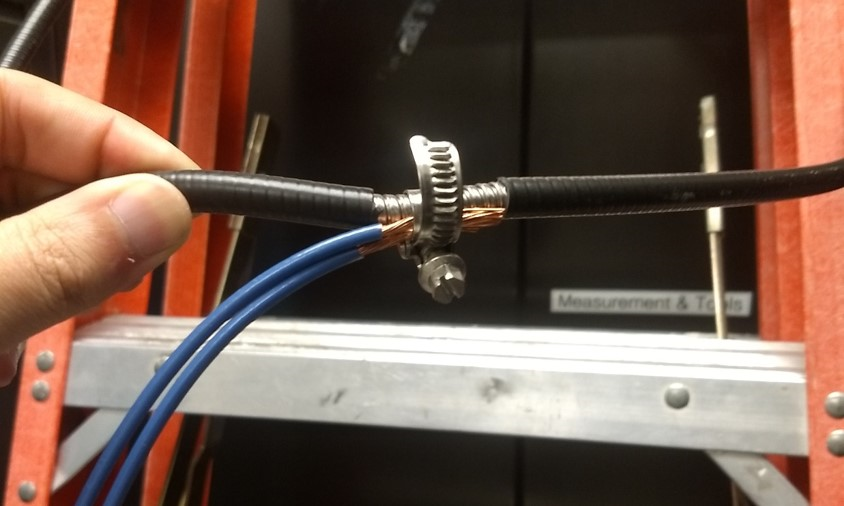}
    \caption{\textit{Top}: WIYN lightning suppression system for the solar feed and the weather station. Cables from outside are captured in lightning isolators and connected to the copper grounding plate; in case of electrical surge, current will be shunted to the grounding plate instead. Data cables (seen emerging from the right side of the isolators) are routed to the USB hub and then transmitted to the NEID server via USB-over-fiber-optic.
    \textit{Bottom}: Lightning grounding for the solar fiber itself. A short section of the outer PVC jacketing was stripped away to allow wire leads to be fastened against the inner steel wrap. These are also connected to the grounding plate as seen in the top panel (blue wires).}
    \label{fig:lightning_ground}
\end{figure}

Due to the danger of lightning strike on Kitt Peak, we were careful to avoid a direct electrical path between the solar feed assembly and any NEID or WIYN electronics, and integrated the solar feed into the existing WIYN lighting suppression system (\autoref{fig:lightning_ground}). Power and data cables from the solar telescope assembly were severed and then re-connected across lightning isolators. While we had already designed the solar fiber with plastic connectors in order to avoid a direct electrical path, we took further precautions by also grounding its steel jacketing directly to the copper grounding plate (\autoref{fig:lightning_ground}, bottom panel). In addition, a new lightning rod---chosen to be significantly taller than the solar feed assembly---was installed nearby on the control building roof to further reduce the chance of a direct strike (\autoref{fig:telescope_installed}, bottom panel).

While all exterior components of the solar feed are designed to be weatherproof, we still wanted a way to safely park the telescope in extreme weather where there is no chance of acquiring useful data. The simplest way to do this is to override the automatic tracking and send the tracker to its home position, pointing due south at zero elevation, where it will stay until manual control is released. The tracker weather-stow must be triggered manually, and generally it is only activated in cases of extremely poor weather. Additionally, we have configured the solar tracker such that it always boots up under manual control. Thus, if observatory power is temporarily lost, the tracker will continue holding at its safe-stow position when power is restored.

Since the installation of the NEID solar feed, it has been subject to adverse weather conditions including, but not limited to: heavy rain, lightning storms, hail, snow, dust storms, and sustained winds of over 50 miles per hour (with gusts exceeding 70~mph). Snow presented a complication since it accumulated on the lip surrounding the lens if the solar telescope was pointing upward, so we elected to send the tracker to its weather-stow safe position whenever snow was forecast. However, we have not observed any damage to the lens, tracker, or any other exterior components of the solar feed in the year since its installation---though the new lightning rod has been struck \emph{at least} once.

\subsection{Path to NEID}
\label{sec:path}

\begin{figure*}[!htbp]
    \centering
    \includegraphics[width=0.8\textwidth]{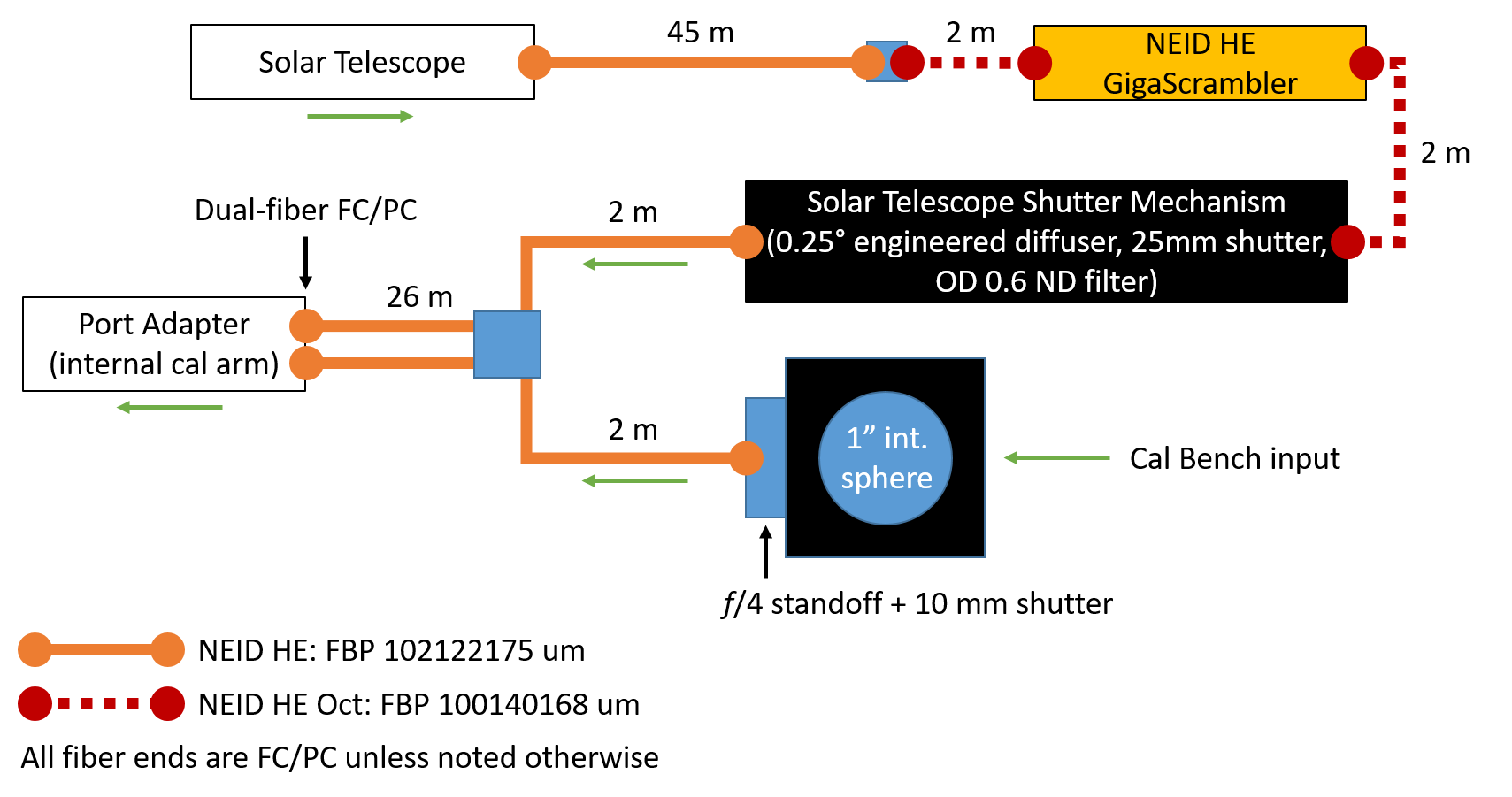}
    \caption{Schematic showing the solar light path up until the NEID port adapter. The solar feed has a dedicated GigaScrambler and an independent shutter mechanism. Like starlight, solar light passes through lengths of octagonal fiber (labeled NEID HE Oct) for additional mode scrambling. The bifurcated port fiber allows either solar light or calibration light to be sent up to the port adapter and then directed into the science fiber.}
    \label{fig:solar_light_path}
\end{figure*}

\autoref{fig:solar_light_path} provides an overview of the solar feed light path. Spatially scrambled sunlight from the solar telescope travels through 45~meters of fiber to reach the NEID calibration room, being routed through the WIYN cable tray along with a fiber-optic cable carrying telemetry from the solar feed assembly sensors. Once they leave the cable tray, these cables are protected in a steel-wrap conduit, which enters the NEID spectrograph room through a small pass-through hole, which is filled with foam to maintain environmental conditions within the spectrograph room.

\begin{figure}[!htbp]
    \centering
    \includegraphics[width=0.45\textwidth]{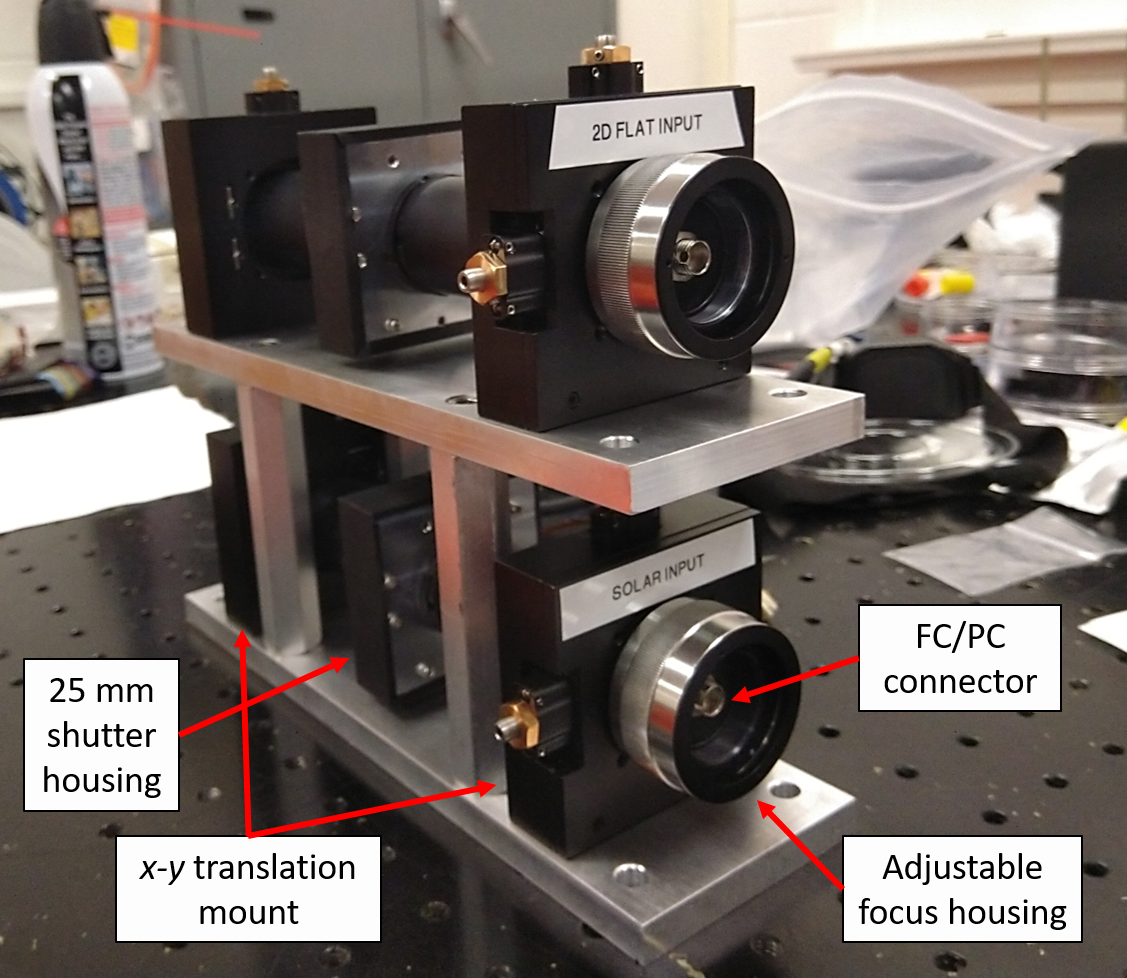}
    \caption{The solar shutter mechanism (bottom tube) and the similar 2D flat field injection tube. The \textit{x-y} translation mounts and adjustable focus housings enable the FC/PC connectors on each end to be moved in all three dimensions and then locked in place. An engineered diffuser, neutral density filter, and a shutter are secured within the tube.}
    \label{fig:solar_shutter}
\end{figure}

Once it reaches the NEID calibration bench, the light is directed through a GigaScrambler (a fiber scrambler, which bends and twists the optical fiber to provide mode scrambling) and the solar shutter mechanism. The solar shutter mechanism (\autoref{fig:solar_shutter}) is a tube which contains, at each end, an achromatic lens doublet and an FC/PC fiber adapter. In the collimated space between the lenses are mounted an RPC Photonics 0.25$\degree$ engineered diffuser, a wedged OD~0.6 neutral density filter,
and a 25~mm shutter. The sunlight exits the fiber and is collimated by the first lens, before being further spatially scrambled by the diffuser and reduced in intensity by the neutral density filter. Then it is refocused into the port fiber at the other end, using two \textit{x-y} translating mounts and adjustable focus housings to center and focus the light on the fiberhead. This shutter mechanism serves as an additional layer of protection to prevent stray solar light entering the port adapter or the rest of the instrument when the solar feed is not in use.

Next, the light is sent up to the NEID port adapter via a bifurcated fiber, which was was custom-made for us by C~Technologies, Inc. It is Y-shaped, with two single steel-jacketed fibers feeding into a junction where the two ``tails'' are combined into a single jacket, with both fibers close to each other in one ferrule at the far end. As depicted in \autoref{fig:solar_light_path}, one tail is attached to the solar shutter mechanism and the other to the calibration bench, allowing either source to be sent up to the internal calibration arm of the NEID port adapter.

\begin{figure*}[!htbp]
    \centering
    \includegraphics[width=0.7\textwidth]{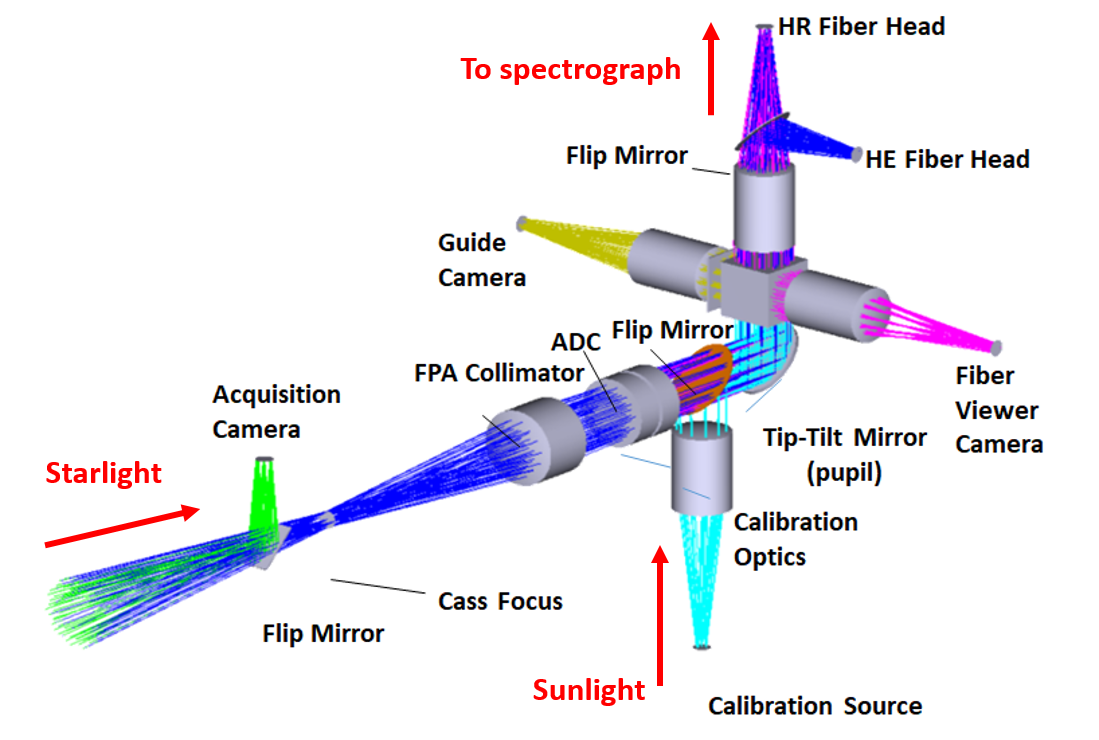}
    \caption{Labeled schematic of the NEID port adapter, adapted from \citet{NEID_Port_Design} and \citet{NEID_Port_Overview}, showing the paths of solar and stellar light. A flip mirror selects between starlight from the WIYN 3.5m or solar light from the internal calibration arm of the port adapter.}
    \label{fig:port_layout}
\end{figure*}

\autoref{fig:port_layout} shows the subsequent light path through the port adapter. This path bypasses the first few optics, which shape the input beam from the WIYN 3.5m Telescope, and adds a lens triplet and a flip mirror in the internal calibration arm. All optics after the atmospheric dispersion corrector (ADC) are identical to the light path traveled by normal science light. A small fraction of the solar light is directed to the Guide Camera and Fiber Viewing Camera by the port beamsplitter; this allows an automated script to adjust \textit{x-y} stages within the port adapter to direct solar light into the fiber head leading to the spectrograph. 

The NEID port adapter is further described in \citet{NEID_Port_Overview} and \citet{NEID_Port_Design}. The overall design of the NEID fiber feed closely resembles that of the Habitable-zone Planet Finder (HPF), mounted on the 10m Hobby-Eberly Telescope---see \citet{Kanodia2018_HPFfibers} for more details.

Early in the design process, we considered feeding solar light through the calibration bench turret as if it were a calibration source. This would have eliminated the need for the bifurcated fiber and stage adjustments in the port adapter, but we decided that this design was unworkable for three major reasons. First, the solar light would have to pass through a second integrating sphere in the calibration bench, resulting in a total loss of $> 10^{12}$; even for the Sun, the resulting photon counts would be too low for any reasonable exposure time. Second, this would have precluded taking solar observations with simultaneous calibration, requiring interleaved calibration frames instead. Finally, this alternate design would bypass the port adapter by sending sunlight directly from the calibration bench to the spectrograph, preventing us from probing the instrumental systematics of the port adapter and the NEID science fiber.

In addition to sunlight, the solar feed assembly also needs to send and receive telemetry from the NEID server located in the calibration room. Due to the distance between the solar feed assembly and the calibration room, all data and communications devices are connected to a USB hub (after passing through the lightning isolation system) and collectively connected to the NEID server via a single fiber-optic cable routed through the WIYN cable tray alongside the solar fiber.

\section{Observations}
\label{sec:observations}

\subsection{Observing Strategy}
\label{sec:observing_strategy}

We have designed our solar observing strategy to maximize the collection of solar data between daily NEID calibration sequences, while also requiring minimal human involvement. We collect $\sim$6 hours of solar data every clear day.

Every morning, when the Sun is 5$\degree$ below the horizon, the solar tracker will automatically move itself to the predicted position of the Sun as computed from its internal model. Once the Sun rises, the tracker will lock on with its active tracking loop and continue following the Sun throughout the day, regardless of whether or not solar data are being taken. 

After the NEID morning calibration sequence is finished, the solar data-collection script will trigger at 16:31 UT (09:31 local time). This script opens the solar shutter mechanism and reconfigures the port adapter, which will center the Sun onto the High Resolution (HR) science fiber, and then takes exposures at a fixed cadence until stopped. If there is not enough flux for the port adapter to properly acquire the Sun on startup, it will try again every 60 seconds until it is successful or the solar script ends. The script will also attempt to re-acquire the Sun after every 60 solar frames to ensure that the source is centered. We take solar data continuously through the daily liquid nitrogen (LN2) fill, which occurs at 17:00 UT (10:00 local). At 22:30 UT (15:30 local), the solar script will automatically stop taking exposures and reset NEID to its normal nighttime observing configuration, so that the afternoon calibration sequence can start on time.

At sunset, the solar tracker will cease tracking once the Sun is 5$\degree$ below the horizon and return to its home position until the next sunrise. In the event of dangerous weather conditions with no hope of solar data, the solar tracker will be manually held at its safe-stow position instead.

We experience significant seasonal variation in solar elevation. The elevation of the Sun at the beginning of daily solar data collection ranges from 20$\degree$ in the winter to 50$\degree$ in the summer; data collection finishes at approximately the same elevation. The maximum solar elevation (at solar noon) ranges from 35$\degree$ in the winter to 82$\degree$ in the summer.

The exposure time for solar frames is 55 seconds. As of 24 August 2021, the gap between solar exposures has been shortened to 28 seconds (previously 38~s), for a total observing cadence of 83 seconds (previously 93~s). The $\sim$1.5 minute cadence is well-suited for our purposes, allowing us to sample the solar p-modes without losing too much of the duty cycle to readout time.

\subsection{Solar Data Processing}
\label{sec:data_reduction}

The solar data are processed by the standard CCF-based NEID Data Reduction Pipeline (DRP), almost identically to the stellar data. This allows us to analyze the solar data as a true reflection of the precision achievable by the NEID instrument and pipeline in combination. For a full description of the NEID DRP, see the documentation on the NEID Data Archive\footnote{\url{https://neid.ipac.caltech.edu/docs/NEID-DRP/}}. However, there are a few crucial differences between the nighttime and daytime pipeline reductions due to the nature of solar observations.

First, the desired ``barycentric correction'' for the Sun is different than that for a normal exoplanet host star. Following \citet{WrightKanodia2020}, we correct for both the barycentric motion of the observer on Earth \emph{and} the reflex motion of the Sun around the barycenter due to the other planets in the Solar System (mainly Jupiter and Saturn).  After applying this correction, any residual variation in the solar RVs should only be from solar activity and instrumental noise. To do this, we have developed additional functionality for \texttt{barycorrpy} \citep{Kanodia2018_barycorrpy} specifically for use with the Sun (and other solar system objects); we use the predictive solar mode, which corrects the velocities as described above. The output of \texttt{barycorrpy} is $z_{\rm{predict}}$, which is then reformatted to resemble a standard barycentric correction: $z_{\rm{B, SS}} = 1 / (1 + z_{\rm{predict}}) - 1$, in order to employ the standard formalism from \citet{WrightEastman2014_barycorr}.

\begin{figure*}[!htbp]
\centering
\includegraphics[width=0.7\textwidth]{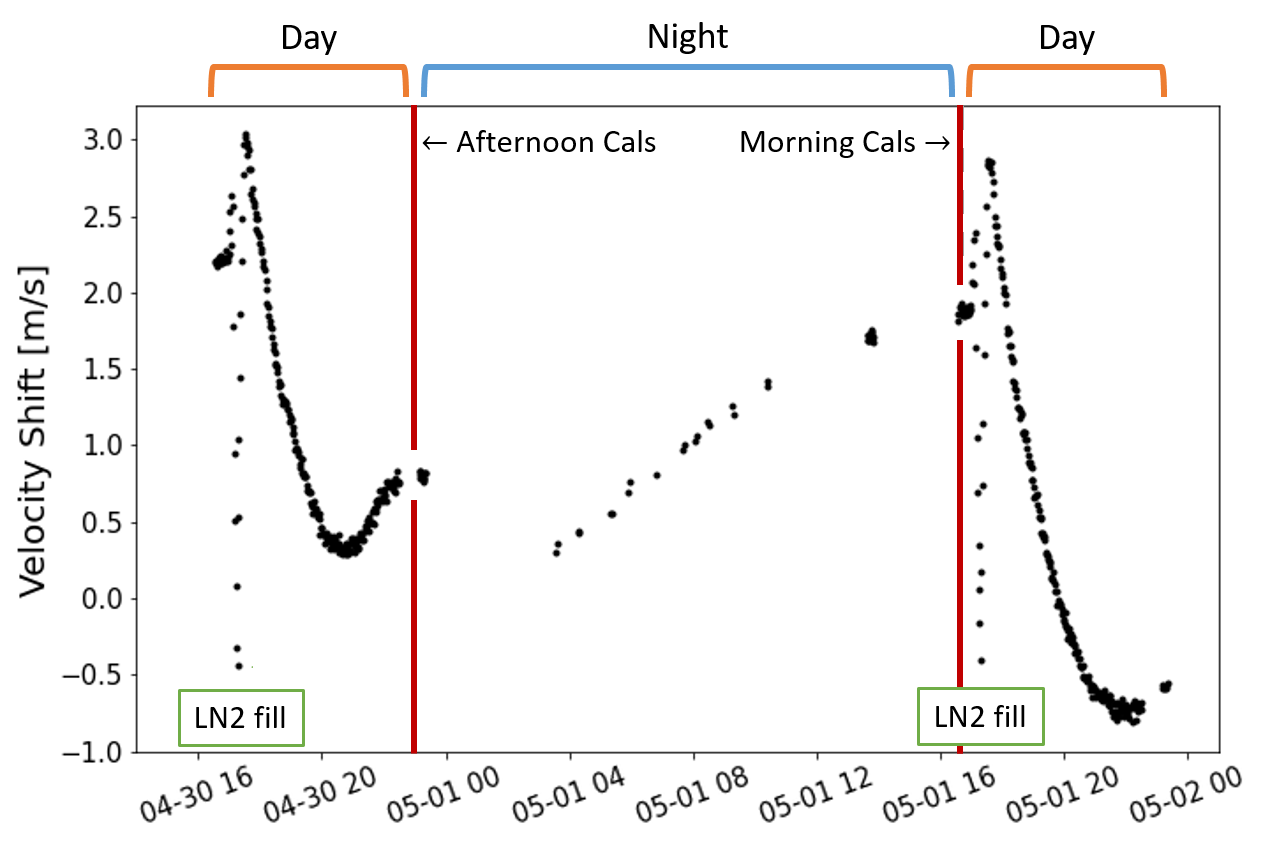}
\caption{Typical daily drift pattern of NEID, traced by the etalon calibration source. The drift pattern on sequential days is similar, but not identical. ``Day'' and ``night'' are bookended by the morning and afternoon calibration sequences (red vertical lines), with solar data being taken over the ``day'' timespan. Drift data is denser during the day because all solar data is taken with simultaneous etalon, while nighttime data may use either simultaneous or bracketed calibrations. Liquid nitrogen (LN2) fills cause a sharp disturbance in the daily drift pattern (marked by green boxes).}
\label{fig:solar_drifts}
\end{figure*}

Second, the instrumental drift patterns are different for solar data as compared to nighttime data. We have observed that the drift slope varies with time since the LN2 fill, and the sudden transient caused by the LN2 fill itself occurs during solar data collection (\autoref{fig:solar_drifts}). Version 1.0 of the NEID DRP uses only the simultaneous etalon data for solar drift correction, and does not use the interpolated drift model used for nighttime data. An update in DRP Version 1.1.2 incorporates an interpolated drift model for the solar data too; this means that the drift is not calculated solely from the simultaneous etalon calibration, but also informed by the data from neighboring frames. All solar data have recently been reprocessed with v1.1.2, and will be reprocessed with any future improvements to the NEID DRP.

One such improvement will be the addition of differential extinction correction. The Sun is a resolved source---half a degree in diameter---and thus there is a slight imbalance in atmospheric extinction between the blueshifted and redshifted sides of the Sun. This leads to a slow RV drift on the order of 1~m/s over the course of a day. \citet{CollierCameron_2019} outline one method of correcting differential extinction, which produces a single (grey) RV offset. However, the true perturbation is a line shape change, and further work will be necessary to determine the effects of this perturbation on different methods for measuring RVs (e.g., CCF-based vs.\ template-matching).

We also note that the CCF mask used for the solar RVs presented in this paper mirrors the ESPRESSO line list. Because ESPRESSO's wavelength coverage does not extend as far into the red as NEID (with its red cutoff at 788~nm, while NEID shows useful spectral orders past its nominal 930~nm red cutoff), we are not currently taking full advantage of the information contained in NEID spectra. In the future, a NEID-specific line list will be developed to encompass the full NEID wavelength range.

\subsection{Current Solar Data}
\label{sec:current_data}

The solar feed assembly was first connected to NEID during spectrograph testing at Penn State in June 2019. Test solar exposures confirmed that it was operational, and it was relocated to the roof above the NEID integration clean room for use throughout the final stability run in August and September. During this period, we were able to collect a few days of solar data, which we used to help quantify instrumental throughput and stability in the lead-up to the pre-ship review. 

The solar telescope was initially installed at WIYN in November 2019, but was not fully operational until January 2020. At that time, port adapter setup and exposures still needed to be triggered manually. In addition to days with poor weather or lack of available NEID personnel at WIYN, solar data was often preempted by daytime port commissioning work and telescope maintenance. However, we were able to acquire short stretches of solar data (up to 2.5~hours) on a handful of days in late January through mid-February of 2020.

Unfortunately, this was followed by a long stretch without solar data. The spectrograph was warmed up and opened for engineering work in late February 2020. Before it had re-stabilized in the subsequent vacuum cycle, Kitt Peak facilities shut down due to COVID-19 in mid-March---meaning that NEID and all its subsystems were powered down to a long-term safe mode. 

\begin{figure*}[!htbp]
    \centering
    \includegraphics[width=\textwidth]{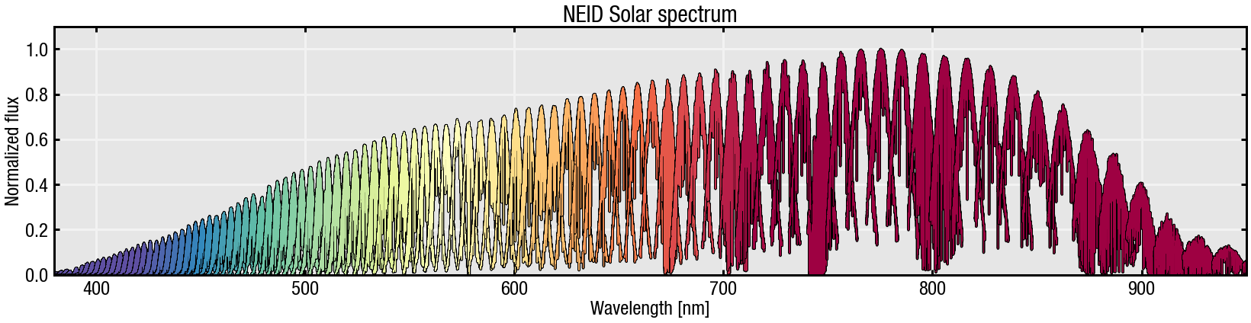}
    \caption{A typical 1D solar spectrum taken with NEID and reduced by the NEID DRP, showing the overall instrumental response and blaze response.}
    \label{fig:solar_spectrum_blazed}
\end{figure*}

The mountain was able to reopen in September 2020 and NEID started cooling again in late October. The solar feed system was brought back up without issue and the outdoor optics were inspected and cleaned. As of December 2020, the solar feed is once again operational and taking solar data on a daily basis. \autoref{fig:solar_spectrum_blazed} shows an example solar spectrum from WIYN.

All NEID solar data can be accessed on the NExScI NEID Solar Archive at \url{https://neid.ipac.caltech.edu/search_solar.php}. These data are available immediately upon being processed by the DRP (typically within 24 hours of being taken), with no proprietary period.

\section{Early RV Results}
\label{sec:early_results}

In the following section, we focus on the daily solar data taken during NEID instrument commissioning at WIYN, from mid-December 2020 to mid-April 2021. While a small amount of solar data was taken prior to this, it is sparsely sampled and the spectrograph had not settled into its current stable operating configuration. Here, we use results from NEID DRP v1.0, as the solar data had not yet been fully reprocessed with DRP v1.1.2 when we performed our analysis.

One of our main goals with the solar feed, beyond that of understanding stellar activity, is to investigate the instrumental stability of NEID. The NEID optical bench is under vacuum and temperature controlled to sub-milli-Kelvin levels \citep{Stefansson_thermal, Robertson_thermal, Robertson_environment}. However, minute temperature fluctuations and mechanical flexure of the bench may still influence the instrumental RV precision. In principle, such effects should be tracked by the calibration fiber, canceling them out from the final reduced RVs. The solar data provides us with a very densely sampled dataset with very high signal-to-noise (SNR), which is extremely useful for investigating instrumental stability.

We present our results below, with the following caveats. The solar feed is an auxiliary component of NEID and did not have the stringent error budget of the spectrograph or the port adapter. Thus, the RV error contributed by its design has not been quantified as rigorously as the other components of NEID. The solar feed also bypasses the WIYN 3.5m Telescope and part of the port adapter, including the tip-tilt loop and the ADCs, and as such its performance may not be reflective of the \emph{entire} NEID system. In addition, we have not accounted for solar RV variability apart from binning our exposures to $\sim$5.6 minutes to approximately average over p-modes; the contribution to RV jitter from granulation and other longer-period effects over our time series are unknown. Therefore, while our solar data cannot tell us the spectrograph's true instrumental RV precision, it can provide an approximate upper bound in the best-case scenario of a dense, high-SNR dataset.

In this early analysis, we filter the solar data aggressively in order to better probe NEID's instrumental precision, rejecting poor or cloudy days of solar observations based on the overall pyrheliometer data for the day. We choose to filter by entire days, as opposed to frame-by-frame, in order to ensure that each day has sufficient data for a daily bin to be meaningful. We discard any days which show frequent and significant drops in flux (a small number of brief cloudy patches on an otherwise clear day is acceptable), because we find a strong correlation between low flux in the pyrheliometer and RV offsets, as a result of uneven obscuration of the solar disk. This is followed by a SNR cut to remove any remaining low-flux points (e.g. occasional clouds). However, we \emph{do not} reject points based on RVs at any stage in this filtering process. Finally, we bin our RVs in sets of 4 exposures (334~s, 5.6 min) to approximately average over the $\sim$5.4-minute solar p-mode period.

\begin{figure*}[!htbp]
    \centering
    \includegraphics[width=\textwidth]{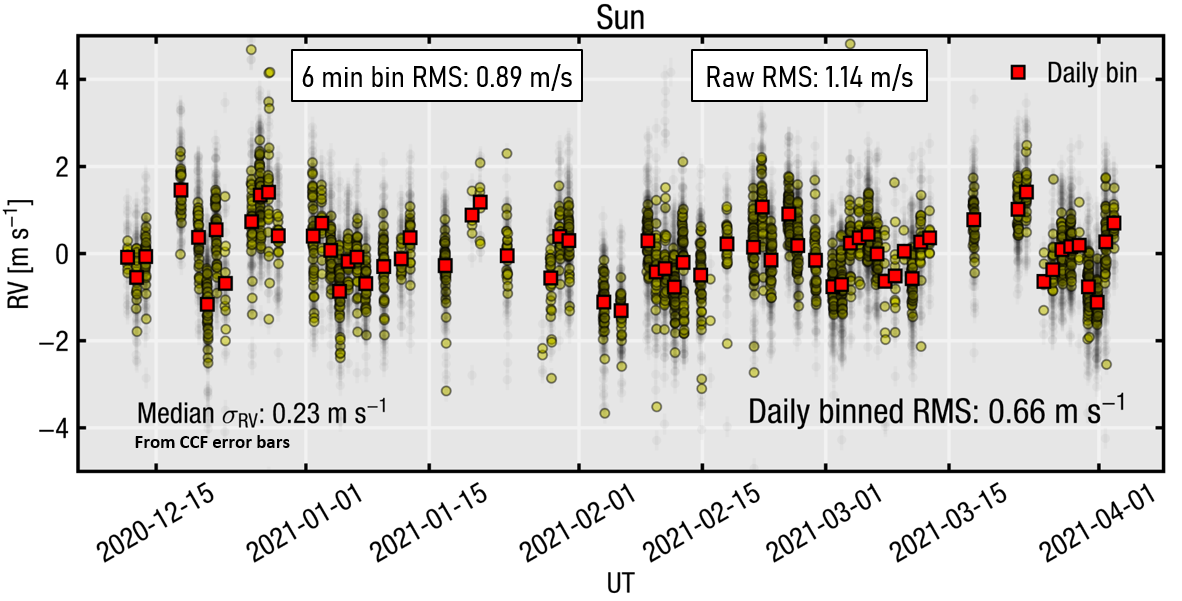}
    \includegraphics[width=\textwidth]{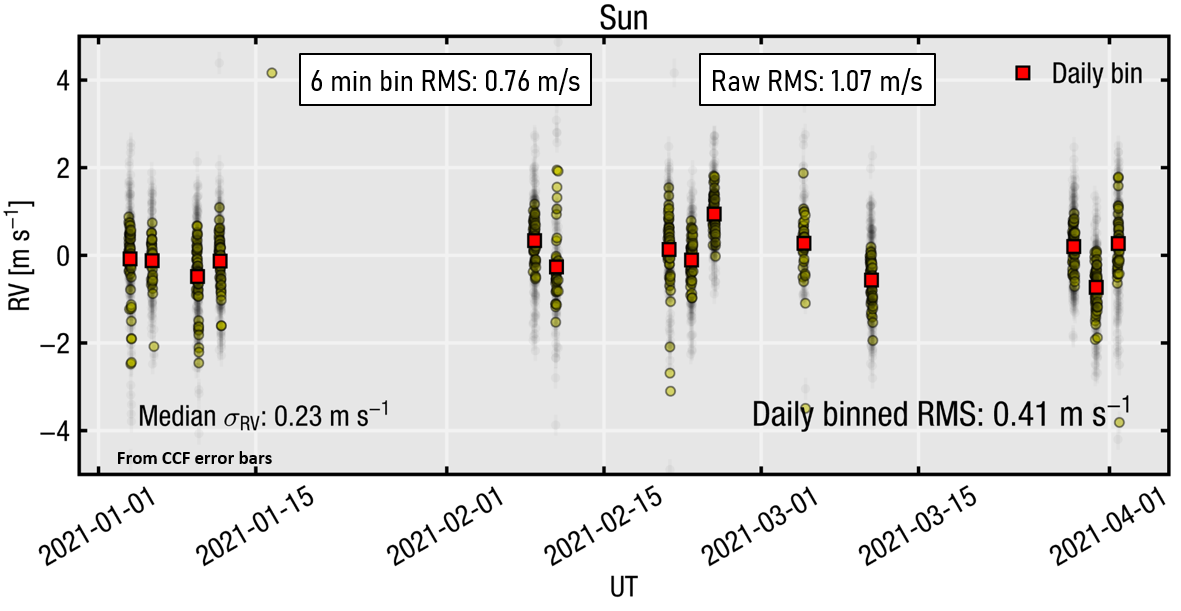}
    \caption{\textit{Top}: NEID solar RVs on all good-quality days during instrument commissioning. The raw RV RMS is 1.14~m/s, which decreases to 0.89~m/s when approximately binning over solar p-modes. Daily binned RMS is 0.66~m/s. \textit{Bottom}: NEID solar RVs on all best-quality days (roughly the clearest 10\% of days) during commissioning. As expected, these RVs perform better than the good-quality days, indicating that even minor cloudiness or solar obscuration has a detrimental effect on the observed RV precision. The raw RMS is 1.07~m/s, which decreases to 0.76~m/s when approximately binning over solar p-modes. Daily binned RMS is 0.41~m/s.}
    \label{fig:rv_good_days}
\end{figure*}

Over 4-5 months of solar data, we show in the top panel of \autoref{fig:rv_good_days} that on the remaining ``good''-quality days, NEID has an RV stability of 1.14~m/s RMS on the Sun. If we bin these observations as described previously, to beat down the p-mode noise, the RMS is reduced to 0.89~m/s, and if we bin over entire days---to probe the long-term instrumental stability of NEID---we achieve a precision of 0.66~m/s.

In an effort to quantify NEID's best-case-scenario precision floor on the Sun, we apply even stricter conditions to the pyrheliometer data. We exclude days which show any sharp flux dips due to clouds, eliminating many of the intermittently cloudy days that were considered ``good'', or which show fluxes consistently discrepant from the smooth curve predicted by the Bird Model (indicating perpetual wispy clouds or haze). We find that this leaves us with the clearest $\sim$10\% of days---which we will refer to as ``best''-quality days---where we find an RV precision of 1.07~m/s (unbinned), 0.76~m/s (5.6-minute bins), and 0.41~m/s (daily bins) over this timespan as shown in the bottom panel of \autoref{fig:rv_good_days}.

\begin{figure*}[!htbp]
    \centering
    \includegraphics[width=0.9\textwidth]{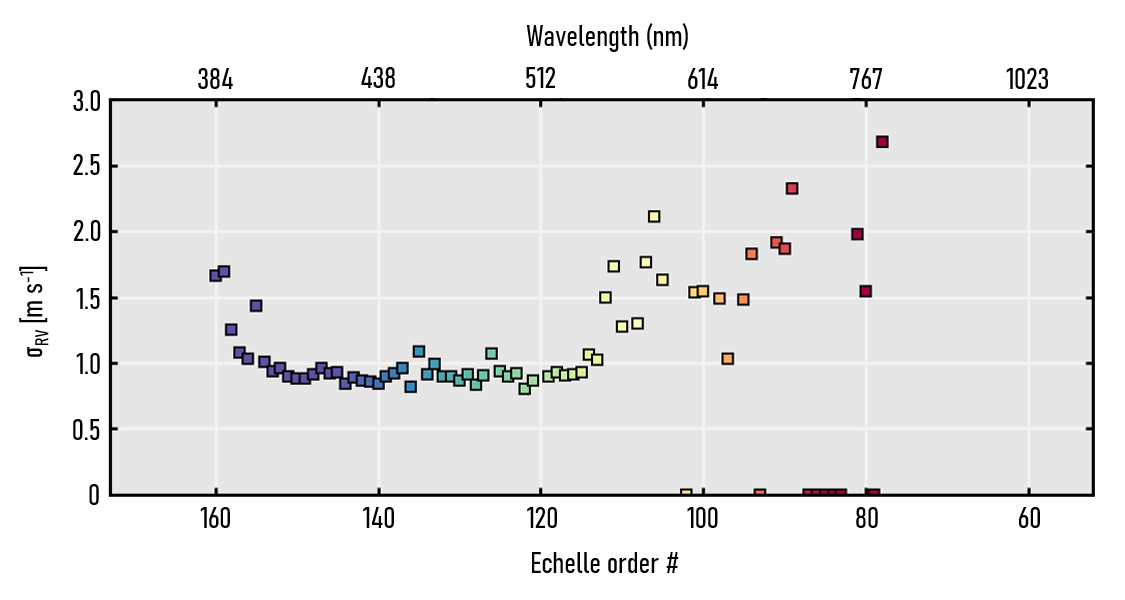}
    \caption{Order-by-order RV precision of NEID on good-quality solar days. Most orders show precisions $\sim$1~m/s when binned over a day, comparable to the cumulative precision of last-generation RV instruments. The decreased precision at the extremes of the wavelength range is not unexpected, as a result of lower overall flux in the bluest orders and a smaller number of RV mask lines in the redder orders, which are most affected by telluric lines. Orders which show $\sigma_{\rm{RV}}$ of zero had no mask lines available to calculate RVs.}
    \label{fig:rv_precision_by_order}
\end{figure*}

\begin{figure*}[!htbp]
    \centering
    \includegraphics[width=\textwidth]{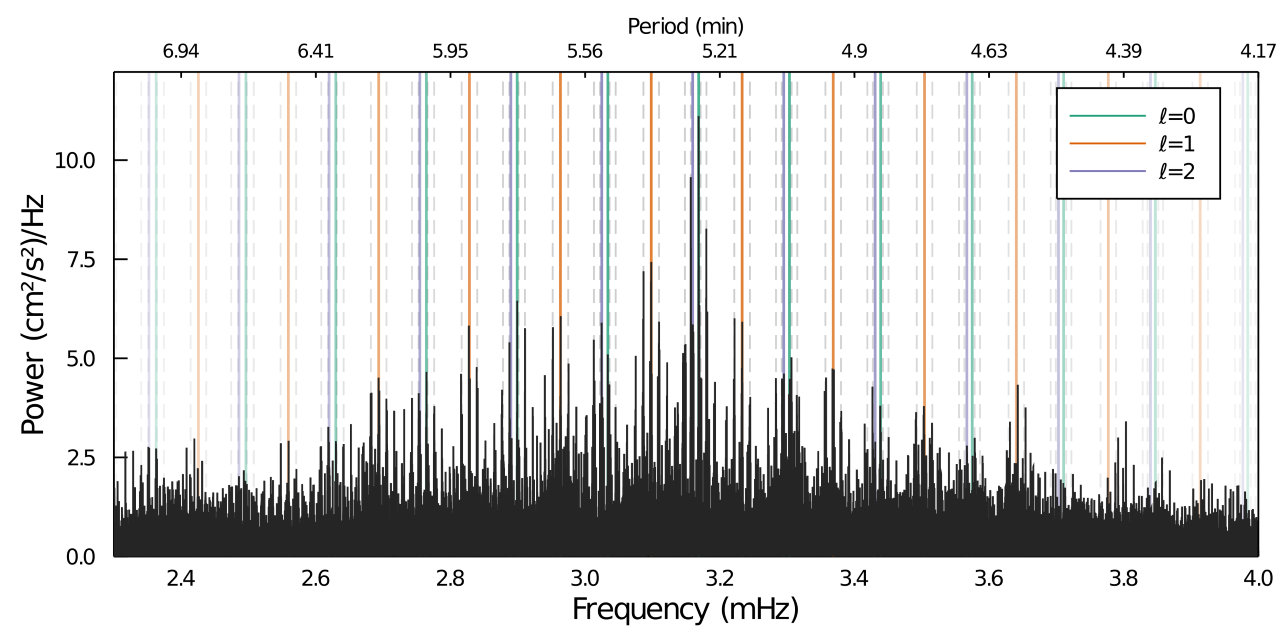}
    \caption{Recovery of solar p-mode frequencies around the $\sim$5.4~min harmonic peak. 
    In order to focus on the oscillation RV signatures, here we have removed points with RVs differing from the overall median RV by more than 5~m/s, and subtracted a linear trend from each day of data before performing two rounds of 3-sigma clipping.
    Known solar p-mode frequencies, determined by BISON \citep{Broomhall2009}, are marked by solid vertical lines ($l$=0 in green, $l$=1 in orange, and $l$=2 in purple). The observed NEID peaks match the BISON frequencies extremely well, differing by median values of 0.42 $\rm{\mu}$Hz, 0.56 $\rm{\mu}$Hz, and 1.5 $\rm{\mu}$Hz for $l$ = 0, 1, and 2, respectively. Dashed vertical lines mark the 1-day aliases of the shown BISON frequencies, which account for the vast majority of additional periodogram peaks.}
    \label{fig:pmodes}
\end{figure*}

In addition, we present preliminary demonstrations of possible future investigations with NEID solar data. \autoref{fig:rv_precision_by_order} depicts the order-by-order RV precision of NEID. The majority of orders are well-behaved with daily solar RV RMS $\sim$1~m/s, offering opportunities for chromatic RV analysis. Furthermore, in \autoref{fig:pmodes}, we recover solar p-mode harmonics around the expected $\sim$5.4-min period peak. Further investigations of stellar activity, while outside the scope of this paper, appear promising, and will be the subject of future publications (e.g., Ervin et al. 2022, in prep; Ford et al. 2022, in prep).

\section{Conclusions and Future Work}
\label{sec:conclusions}

We have designed and built a solar feed system for NEID, the data from which have already allowed us to demonstrate our instrumental precision. With carefully considered hardware choices, we were able to construct a durable and reliable system while meeting our aggressive development timeline. Data collected by the solar telescope allows us to place approximate upper bounds on the NEID instrumental RV stability (daily bins) of 0.66~m/s under ``good'' conditions, and 0.41~m/s under ``best'' conditions. Our solar observations have also enabled us to monitor long-term instrumental drift and resolve solar p-modes. 

Future observations with the NEID solar feed will enable us to continue our investigations into instrument systematics and RV pipeline improvements, but will also serve far broader purposes. A large and homogeneous RV dataset of the Sun offers many avenues for tackling the problem of RV jitter due to stellar activity. High-cadence, long-baseline observations allow us to test techniques for minimizing the impact of pulsation, granulation, and longer-term RV variations. Spectral activity indicators and new metrics for quantitatively linking activity to RV variations can be developed in a situation where we can verify our findings. Such rich datasets also provide an ideal test case for advanced statistical modeling of stellar jitter, and novel methods of deriving stellar RVs that allow us to take advantage of a larger fraction of the information encoded in a stellar spectrum. 

In addition, such solar feed systems on precision RV machines open up the possibility of simultaneous monitoring of solar RVs with multiple instruments. Ultimately, we hope that new techniques developed with the solar feed data allow us to overcome the stellar activity barrier and reach the $<$ 10~cm/s precision necessary to find Earth-mass planets in or near Habitable Zones around the nearest stars, and take the next step toward spectroscopy and characterization of these planets.

\section*{Acknowledgments}

Based on observations at Kitt Peak National Observatory, National Optical Astronomy Observatory, which is operated by the Association of Universities for Research in Astronomy (AURA) under a cooperative agreement with the National Science Foundation. 
 Data presented were obtained by the NEID spectrograph built by Penn State University and operated at the WIYN Observatory by NOIRLab, under the NN-EXPLORE partnership of the National Aeronautics and Space Administration and the National Science Foundation. WIYN is a joint facility of the University of Wisconsin–Madison, Indiana University, NSF’s NOIRLab, the Pennsylvania State University, Purdue University, University of California, Irvine, and the University of Missouri. The authors are honored to be permitted to conduct astronomical research on Iolkam Du'ag (Kitt Peak), a mountain with particular significance to the Tohono O'odham.

This research was supported by Heising-Simons Foundation Grant \# 2019-1177.
This work was partially supported by funding from the Center for Exoplanets and Habitable Worlds. The Center for Exoplanets and Habitable Worlds and the Penn State Extraterrestrial Intelligence Center are supported by the Pennsylvania State University and the Eberly College of Science.
CIC acknowledges support by NASA Headquarters under the NASA Earth and Space Science Fellowship Program through grant 80NSSC18K1114.
Part of this work was performed for the Jet Propulsion Laboratory, California Institute of Technology, sponsored by the United States Government under the Prime Contract 80NM0018D0004 between Caltech and NASA.

Computations for this research were performed on the Pennsylvania State University’s Institute for Computational and Data Sciences’ Roar supercomputer.
Processing of data included here used an allocation of computer time from the Research Computing High Performance Computing cluster at the University of Arizona, which is gratefully acknowledged.
We acknowledge BJ Fulton and colleagues at the NASA Exoplanet Science Institute (NExScI) for their efforts in hosting the NEID Solar Data Archive.

This research has made use of NASA's Astrophysics Data System Bibliographic Services.
This research made use of Astropy (\url{http://www.astropy.org}), a community-developed core Python package for Astronomy \cite{AstropyCollaboration2013_astropy, AstropyCollaboration2018_astropy}.

\facilities{NEID/WIYN 3.5m}
\software{
\texttt{astropy} \citep{AstropyCollaboration2018_astropy},
\texttt{barycorrpy} \citep{Kanodia2018_barycorrpy}, 
\texttt{matplotlib} \citep{Hunter2007_matplotlib},
\texttt{numpy} \citep{vanderWalt2011_numpy},
\texttt{pandas} \citep{McKinney2010_pandas},
\texttt{pvlib} \citep{Holmgren2018_pvlib},
\texttt{scipy} \citep{Virtanen2020_scipy}
}

\appendix

We list here in \autoref{tab:parts_table} the major components of the NEID solar feed and their costs.

\movetabledown=5cm
\begin{rotatetable*}
\begin{deluxetable*}{llllll}
\tabletypesize{\footnotesize}
\tablecaption{List of major components for the NEID solar feed, grouped by subsystem: solar tracker, optics assembly, sensors, shutter mechanism, and fiber optics. \label{tab:parts_table}}
\centering
\startdata
\\ \hline
\textbf{Vendor}                   & \textbf{Part No.}            & \textbf{Part Name}                                                               & \textbf{Quantity} & \textbf{Cost Ea.} & \textbf{Cost Total}      \\
\hline
EKO Instruments          & STR-22G-S-DC-SET-10 & Double arm sun tracker (w/10m DC power cable and 3m comm cable)         & 1        & 12,331.00  & 12,331.00 \\
EKO Instruments          & MS-57               & Pyrheliometer (w/CR-57 clamping ring and 10m signal cable)              & 1        & 3,287.00   & 3,287.00  \\
EKO Instruments          & STR-22-2A           & Dual pyrheliometer mount (secondary side)                               & 1        & 2,375.00   & 2,375.00  \\
\hline
Edmund Optics            & 88-596-INK          & 75mm dia. 200mm FL achromatic doublet (VIS-NIR, edge-blackened)         & 1        & 309.00     & 309.00    \\
Thorlabs                 & IS200               & 2" integrating sphere (no sensor, 3 ports)                              & 1        & 1,069.51   & 1,069.51  \\
Thorlabs                 & SM05CP2C            & 0.5" integrating sphere end cap                                         & 2        & 30.30      & 60.60     \\
Thorlabs                 & SM05FC              & 0.5" FC/PC adapter with external SM05 threads                           & 1        & 26.52      & 26.52     \\
Thorlabs                 & SM1A1               & Adapter with external SM05 threads  and internal SM1 threads            & 1        & 21.86      & 21.86     \\
Thorlabs                 & SM05L03             & SM05 lens tube, 0.3" length                                             & 2        & 14.21      & 28.42     \\
Thorlabs                 & SM05L05             & SM05 lens tube, 0.5" length                                             & 1        & 14.52      & 14.52     \\
Thorlabs                 & SM05L10             & SM05 lens tube, 1" length                                               & 2        & 15.55      & 31.10     \\
Thorlabs                 & SM1L05              & SM1 lens tube, 0.5" length                                              & 1        & 12.97      & 12.97     \\
McMaster-Carr            & 3006T690            & 304 stainless steel hanger (w/closure bolt and nut)                     & 1        & 6.81       & 6.81      \\
McMaster-Carr            & 7119K730            & 3/4" liquid-tight flexible metal conduit adapter with locknut           & 1        & 3.32       & 3.32      \\
McMaster-Carr            & 9559K136            & Dash number 040 ultra-chemical-resistant rigid PTFE O-ring              & 2        & 3.74       & 7.48      \\
McMaster-Carr            & 9557K131            & Dash number 040 water- and steam-resistant EPDM O-ring (pack of 25)     & 1        & 8.98       & 8.98      \\
McMaster-Carr            & 9560K430            & Dash number 016 ultra-chemical-resistant PTFE backup ring (pack of 50)  & 1        & 11.46      & 11.46     \\
McMaster-Carr            & 8609K48             & 1/16" weather-resistant airtight butyl rubber adhesive-back strips      & 1        & 12.31      & 12.31     \\
SkyGeek                  & ProtekAir 8415      & 18mm dehydrator spark plug                                              & 3        & 12.60      & 37.80     \\
Hilltop Tech. Lab. & Custom              & Housing for solar telescope                                             & 1        & 2,282.11   & 2,282.11  \\
\hline
Dracal                   & USB-PTH200          & USB atmospheric, temperature, and relative humidity sensor              & 1        & 74.31      & 74.31     \\
LabJack                  & U6-PRO              & USB multifunction DAQ                                                   & 1        & 410.00     & 410.00    \\
\hline
Edmund Optics            & 49-355              & 25mm dia. 45mm FL achromatic doublet (VIS-NIR)                          & 2        & 100.00     & 200.00    \\
Thorlabs                 & ST1XY-A             & XY translation mount                                                    & 2        & 363.59     & 727.18    \\
Thorlabs                 & SM1ZM               & Zoom housing for 1" optics                                              & 2        & 176.39     & 352.78    \\
Thorlabs                 & SM1FC               & FC/PC adapter                                                           & 2        & 31.38      & 62.76     \\
Thorlabs                 & SM1L20              & 2" SM1 lens tube                                                        & 2        & 17.00      & 34.00     \\
Thorlabs                 & NDW06B              & Wedged reflective 25mm ND filter (OD 0.6)                               & 1        & 55.19      & 55.19     \\
RPC Photonics            & EDC-0.25-17632-A    & Engineered diffuser (0.25 degree, 1" round)                             & 1        & 320.00     & 320.00    \\
Vincent Associates       & DSS25B1T0           & 25mm bi-stable shutter                                                  & 1        & 445.00     & 445.00    \\
\hline
Thorlabs                 & Custom              & Fiber optic patch cable (FG105ACA w/FT061PS tubing, FC/PC - FC/PC, 40m) & 1        & 872.16     & 872.16    \\
FiberTech Optica         & Custom              & Fiber optic patch cable (FBP102122145, FC/PC - FC/PC, 45m)              & 2        & 1,436.00   & 2,872.00  \\
\\ \hline
\textbf{TOTAL}                    &                     &                                                                         &          &            & \textbf{28,363.15} \\
\enddata
\end{deluxetable*}
\end{rotatetable*}

\bibliography{refs, neid_refs, software_refs}

\begin{thebibliography}{}
\expandafter\ifx\csname natexlab\endcsname\relax\def\natexlab#1{#1}\fi
\providecommand{\url}[1]{\href{#1}{#1}}
\providecommand{\dodoi}[1]{doi:~\href{http://doi.org/#1}{\nolinkurl{#1}}}
\providecommand{\doeprint}[1]{\href{http://ascl.net/#1}{\nolinkurl{http://ascl.net/#1}}}
\providecommand{\doarXiv}[1]{\href{https://arxiv.org/abs/#1}{\nolinkurl{https://arxiv.org/abs/#1}}}

\bibitem[{{Astropy Collaboration} {et~al.}(2013){Astropy Collaboration},
  {Robitaille}, {Tollerud}, {Greenfield}, {Droettboom}, {Bray}, {Aldcroft},
  {Davis}, {Ginsburg}, {Price-Whelan}, {Kerzendorf}, {Conley}, {Crighton},
  {Barbary}, {Muna}, {Ferguson}, {Grollier}, {Parikh}, {Nair}, {Unther},
  {Deil}, {Woillez}, {Conseil}, {Kramer}, {Turner}, {Singer}, {Fox}, {Weaver},
  {Zabalza}, {Edwards}, {Azalee Bostroem}, {Burke}, {Casey}, {Crawford},
  {Dencheva}, {Ely}, {Jenness}, {Labrie}, {Lim}, {Pierfederici}, {Pontzen},
  {Ptak}, {Refsdal}, {Servillat}, \&
  {Streicher}}]{AstropyCollaboration2013_astropy}
{Astropy Collaboration}, {Robitaille}, T.~P., {Tollerud}, E.~J., {et~al.} 2013,
  \aap, 558, A33, \dodoi{10.1051/0004-6361/201322068}

\bibitem[{{Astropy Collaboration} {et~al.}(2018){Astropy Collaboration},
  {Price-Whelan}, {Sip{\H{o}}cz}, {G{\"u}nther}, {Lim}, {Crawford}, {Conseil},
  {Shupe}, {Craig}, {Dencheva}, {Ginsburg}, {Vand erPlas}, {Bradley},
  {P{\'e}rez-Su{\'a}rez}, {de Val-Borro}, {Aldcroft}, {Cruz}, {Robitaille},
  {Tollerud}, {Ardelean}, {Babej}, {Bach}, {Bachetti}, {Bakanov}, {Bamford},
  {Barentsen}, {Barmby}, {Baumbach}, {Berry}, {Biscani}, {Boquien}, {Bostroem},
  {Bouma}, {Brammer}, {Bray}, {Breytenbach}, {Buddelmeijer}, {Burke},
  {Calderone}, {Cano Rodr{\'\i}guez}, {Cara}, {Cardoso}, {Cheedella}, {Copin},
  {Corrales}, {Crichton}, {D'Avella}, {Deil}, {Depagne}, {Dietrich}, {Donath},
  {Droettboom}, {Earl}, {Erben}, {Fabbro}, {Ferreira}, {Finethy}, {Fox},
  {Garrison}, {Gibbons}, {Goldstein}, {Gommers}, {Greco}, {Greenfield},
  {Groener}, {Grollier}, {Hagen}, {Hirst}, {Homeier}, {Horton}, {Hosseinzadeh},
  {Hu}, {Hunkeler}, {Ivezi{\'c}}, {Jain}, {Jenness}, {Kanarek}, {Kendrew},
  {Kern}, {Kerzendorf}, {Khvalko}, {King}, {Kirkby}, {Kulkarni}, {Kumar},
  {Lee}, {Lenz}, {Littlefair}, {Ma}, {Macleod}, {Mastropietro}, {McCully},
  {Montagnac}, {Morris}, {Mueller}, {Mumford}, {Muna}, {Murphy}, {Nelson},
  {Nguyen}, {Ninan}, {N{\"o}the}, {Ogaz}, {Oh}, {Parejko}, {Parley}, {Pascual},
  {Patil}, {Patil}, {Plunkett}, {Prochaska}, {Rastogi}, {Reddy Janga},
  {Sabater}, {Sakurikar}, {Seifert}, {Sherbert}, {Sherwood-Taylor}, {Shih},
  {Sick}, {Silbiger}, {Singanamalla}, {Singer}, {Sladen}, {Sooley},
  {Sornarajah}, {Streicher}, {Teuben}, {Thomas}, {Tremblay}, {Turner},
  {Terr{\'o}n}, {van Kerkwijk}, {de la Vega}, {Watkins}, {Weaver}, {Whitmore},
  {Woillez}, {Zabalza}, \& {Astropy
  Contributors}}]{AstropyCollaboration2018_astropy}
{Astropy Collaboration}, {Price-Whelan}, A.~M., {Sip{\H{o}}cz}, B.~M., {et~al.}
  2018, \aj, 156, 123, \dodoi{10.3847/1538-3881/aabc4f}

\bibitem[{Bird \& Hulstrom(1981)}]{Bird_1981}
Bird, R., \& Hulstrom, R. 1981, Simplified clear sky model for direct and
  diffuse insolation on horizontal surfaces, Tech. Rep. SERI/TR-642-761, Solar
  Energy Research Institute, Golden, CO, \dodoi{10.2172/6510849}

\bibitem[{{Broomhall} {et~al.}(2009){Broomhall}, {Chaplin}, {Davies},
  {Elsworth}, {Fletcher}, {Hale}, {Miller}, \& {New}}]{Broomhall2009}
{Broomhall}, A.~M., {Chaplin}, W.~J., {Davies}, G.~R., {et~al.} 2009, \mnras,
  396, L100, \dodoi{10.1111/j.1745-3933.2009.00672.x}

\bibitem[{{Chaplin} {et~al.}(2019){Chaplin}, {Howe}, {Basu}, {Elsworth},
  {Milbourne}, {Haywood}, {Davies}, {Hale}, {Miglio}, \& {Ross}}]{Chaplin_2019}
{Chaplin}, W.~J., {Howe}, R., {Basu}, S., {et~al.} 2019, \mnras, 489, L86,
  \dodoi{10.1093/mnrasl/slz132}

\bibitem[{{Claudi} {et~al.}(2018){Claudi}, {Ghedina}, {Pace}, {Gallorini}, {Di
  Giorgio}, {Liu}, {Tozzi}, {Lanza}, {Micela}, {Molinari}, {Phillips}, \&
  {Tripodo}}]{Claudi_2018}
{Claudi}, R., {Ghedina}, A., {Pace}, E., {et~al.} 2018, in Society of
  Photo-Optical Instrumentation Engineers (SPIE) Conference Series, Vol. 10700,
  \procspie, 107004N, \dodoi{10.1117/12.2312458}

\bibitem[{{Collier Cameron} {et~al.}(2019){Collier Cameron}, {Mortier},
  {Phillips}, {Dumusque}, {Haywood}, {Langellier}, {Watson}, {Cegla}, {Costes},
  {Charbonneau}, {Coffinet}, {Latham}, {Lopez-Morales}, {Malavolta},
  {Maldonado}, {Micela}, {Milbourne}, {Molinari}, {Saar}, {Thompson},
  {Buchschacher}, {Cecconi}, {Cosentino}, {Ghedina}, {Glenday}, {Gonzalez},
  {Li}, {Lodi}, {Lovis}, {Pepe}, {Poretti}, {Rice}, {Sasselov}, {Sozzetti},
  {Szentgyorgyi}, {Udry}, \& {Walsworth}}]{CollierCameron_2019}
{Collier Cameron}, A., {Mortier}, A., {Phillips}, D., {et~al.} 2019, \mnras,
  487, 1082, \dodoi{10.1093/mnras/stz1215}

\bibitem[{{Collier Cameron} {et~al.}(2020){Collier Cameron}, {Ford}, {Shahaf},
  {Aigrain}, {Dumusque}, {Haywood}, {Mortier}, {Phillips}, {Buchhave},
  {Cecconi}, {Cegla}, {Cosentino}, {Cretignier}, {Ghedina}, {Gonzalez},
  {Latham}, {Lodi}, {Lopez-Morales}, {Micela}, {Molinari}, {Pepe}, {Piotto},
  {Poretti}, {Queloz}, {San Juan}, {Segransan}, {Sozzetti}, {Szentgyorgyi},
  {Thompson}, {Udry}, \& {Watson}}]{CollierCameron_2020}
{Collier Cameron}, A., {Ford}, E.~B., {Shahaf}, S., {et~al.} 2020, arXiv
  e-prints, arXiv:2011.00018.
\newblock \doarXiv{2011.00018}

\bibitem[{{Crass} {et~al.}(2021){Crass}, {Gaudi}, {Leifer}, {Beichman},
  {Bender}, {Blackwood}, {Burt}, {Callas}, {Cegla}, {Diddams}, {Dumusque},
  {Eastman}, {Ford}, {Fulton}, {Gibson}, {Halverson}, {Haywood}, {Hearty},
  {Howard}, {Latham}, {L{\"o}hner-B{\"o}ttcher}, {Mamajek}, {Mortier},
  {Newman}, {Plavchan}, {Quirrenbach}, {Reiners}, {Robertson}, {Roy}, {Schwab},
  {Seifahrt}, {Szentgyorgyi}, {Terrien}, {Teske}, {Thompson}, \&
  {Vasisht}}]{Crass2021}
{Crass}, J., {Gaudi}, B.~S., {Leifer}, S., {et~al.} 2021, arXiv e-prints,
  arXiv:2107.14291.
\newblock \doarXiv{2107.14291}

\bibitem[{{Cretignier} {et~al.}(2020){Cretignier}, {Dumusque}, {Allart},
  {Pepe}, \& {Lovis}}]{Cretignier_2020}
{Cretignier}, M., {Dumusque}, X., {Allart}, R., {Pepe}, F., \& {Lovis}, C.
  2020, \aap, 633, A76, \dodoi{10.1051/0004-6361/201936548}

\bibitem[{{de Beurs} {et~al.}(2020){de Beurs}, {Vanderburg}, {Shallue},
  {Dumusque}, {Collier Cameron}, {Buchhave}, {Cosentino}, {Ghedina}, {Haywood},
  {Langellier}, {Latham}, {L{\'o}pez-Morales}, {Mayor}, {Micela}, {Milbourne},
  {Mortier}, {Molinari}, {Pepe}, {Phillips}, {Pinamonti}, {Piotto}, {Rice},
  {Sasselov}, {Sozzetti}, {Udry}, \& {Watson}}]{deBeurs_2020}
{de Beurs}, Z.~L., {Vanderburg}, A., {Shallue}, C.~J., {et~al.} 2020, arXiv
  e-prints, arXiv:2011.00003.
\newblock \doarXiv{2011.00003}

\bibitem[{{Dumusque}(2018)}]{Dumusque_2018}
{Dumusque}, X. 2018, \aap, 620, A47, \dodoi{10.1051/0004-6361/201833795}

\bibitem[{{Dumusque} {et~al.}(2011){Dumusque}, {Udry}, {Lovis}, {Santos}, \&
  {Monteiro}}]{Dumusque_2011}
{Dumusque}, X., {Udry}, S., {Lovis}, C., {Santos}, N.~C., \& {Monteiro},
  M.~J.~P.~F.~G. 2011, \aap, 525, A140, \dodoi{10.1051/0004-6361/201014097}

\bibitem[{{Dumusque} {et~al.}(2015){Dumusque}, {Glenday}, {Phillips},
  {Buchschacher}, {Collier Cameron}, {Cecconi}, {Charbonneau}, {Cosentino},
  {Ghedina}, {Latham}, {Li}, {Lodi}, {Lovis}, {Molinari}, {Pepe}, {Udry},
  {Sasselov}, {Szentgyorgyi}, \& {Walsworth}}]{Dumusque_2015}
{Dumusque}, X., {Glenday}, A., {Phillips}, D.~F., {et~al.} 2015, \apjl, 814,
  L21, \dodoi{10.1088/2041-8205/814/2/L21}

\bibitem[{{Dumusque} {et~al.}(2020){Dumusque}, {Cretignier}, {Sosnowska},
  {Buchschacher}, {Lovis}, {Phillips}, {Pepe}, {Alesina}, {Buchhave},
  {Burnier}, {Cecconi}, {Cegla}, {Cloutier}, {Collier Cameron}, {Cosentino},
  {Ghedina}, {Gonzalez}, {Haywood}, {Latham}, {Lodi}, {Lopez-Morales},
  {Maldonado}, {Malavolta}, {Micela}, {Molinari}, {Mortier}, {Perez Ventura},
  {Pinamonti}, {Poretti}, {Rice}, {Riverol}, {Riverol}, {San Juan},
  {Segransan}, {Sozzetti}, {Thompson}, {Udry}, \& {Wilson}}]{Dumusque_2020}
{Dumusque}, X., {Cretignier}, M., {Sosnowska}, D., {et~al.} 2020, arXiv
  e-prints, arXiv:2009.01945.
\newblock \doarXiv{2009.01945}

\bibitem[{{EKO Instruments}(2019)}]{EKO_antarctica}
{EKO Instruments}. 2019, EKO Contributes to Antarctic Weather Observation
  Project,
  \url{https://eko-eu.com/news/2019-02-12-eko-contributes-to-antarctic-weather-observation-project}

\bibitem[{{ESO}(2018)}]{HARPS_HELIOS}
{ESO}. 2018, {HARPS Sees Sunshine} for the First Time,
  \url{https://www.eso.org/public/usa/announcements/ann18033/#1}

\bibitem[{{Fischer} {et~al.}(2016){Fischer}, {Anglada-Escude}, {Arriagada},
  {Baluev}, {Bean}, {Bouchy}, {Buchhave}, {Carroll}, {Chakraborty}, {Crepp},
  {Dawson}, {Diddams}, {Dumusque}, {Eastman}, {Endl}, {Figueira}, {Ford},
  {Foreman-Mackey}, {Fournier}, {F{\H{u}}r{\'e}sz}, {Gaudi}, {Gregory},
  {Grundahl}, {Hatzes}, {H{\'e}brard}, {Herrero}, {Hogg}, {Howard}, {Johnson},
  {Jorden}, {Jurgenson}, {Latham}, {Laughlin}, {Loredo}, {Lovis}, {Mahadevan},
  {McCracken}, {Pepe}, {Perez}, {Phillips}, {Plavchan}, {Prato}, {Quirrenbach},
  {Reiners}, {Robertson}, {Santos}, {Sawyer}, {Segransan}, {Sozzetti},
  {Steinmetz}, {Szentgyorgyi}, {Udry}, {Valenti}, {Wang}, {Wittenmyer}, \&
  {Wright}}]{Fischer_2016}
{Fischer}, D.~A., {Anglada-Escude}, G., {Arriagada}, P., {et~al.} 2016, \pasp,
  128, 066001, \dodoi{10.1088/1538-3873/128/964/066001}

\bibitem[{{Gibson} {et~al.}(2016){Gibson}, {Howard}, {Marcy}, {Edelstein},
  {Wishnow}, \& {Poppett}}]{KPF_main}
{Gibson}, S.~R., {Howard}, A.~W., {Marcy}, G.~W., {et~al.} 2016, in Society of
  Photo-Optical Instrumentation Engineers (SPIE) Conference Series, Vol. 9908,
  Ground-based and Airborne Instrumentation for Astronomy VI, ed. C.~J.
  {Evans}, L.~{Simard}, \& H.~{Takami}, 990870, \dodoi{10.1117/12.2233334}

\bibitem[{{Halverson} {et~al.}(2016){Halverson}, {Terrien}, {Mahadevan}, {Roy},
  {Bender}, {Stef{\'a}nsson}, {Monson}, {Levi}, {Hearty}, {Blake}, {McElwain},
  {Schwab}, {Ramsey}, {Wright}, {Wang}, {Gong}, \& {Roberston}}]{NEID_budget}
{Halverson}, S., {Terrien}, R., {Mahadevan}, S., {et~al.} 2016, in \procspie,
  Vol. 9908, Ground-based and Airborne Instrumentation for Astronomy VI,
  99086P, \dodoi{10.1117/12.2232761}

\bibitem[{{Haywood} {et~al.}(2014){Haywood}, {Collier Cameron}, {Queloz},
  {Barros}, {Deleuil}, {Fares}, {Gillon}, {Lanza}, {Lovis}, {Moutou}, {Pepe},
  {Pollacco}, {Santerne}, {S{\'e}gransan}, \& {Unruh}}]{Haywood_2014}
{Haywood}, R.~D., {Collier Cameron}, A., {Queloz}, D., {et~al.} 2014, \mnras,
  443, 2517, \dodoi{10.1093/mnras/stu1320}

\bibitem[{{Haywood} {et~al.}(2016){Haywood}, {Collier Cameron}, {Unruh},
  {Lovis}, {Lanza}, {Llama}, {Deleuil}, {Fares}, {Gillon}, {Moutou}, {Pepe},
  {Pollacco}, {Queloz}, \& {S{\'e}gransan}}]{Haywood_2016}
{Haywood}, R.~D., {Collier Cameron}, A., {Unruh}, Y.~C., {et~al.} 2016, \mnras,
  457, 3637, \dodoi{10.1093/mnras/stw187}

\bibitem[{{Haywood} {et~al.}(2020){Haywood}, {Milbourne}, {Saar}, {Mortier},
  {Phillips}, {Charbonneau}, {Collier Cameron}, {Cegla}, {Meunier}, \&
  {Palumbo}}]{Haywood_2020}
{Haywood}, R.~D., {Milbourne}, T.~W., {Saar}, S.~H., {et~al.} 2020, arXiv
  e-prints, arXiv:2005.13386.
\newblock \doarXiv{2005.13386}

\bibitem[{{Holmgren} {et~al.}(2018){Holmgren}, {Hansen}, \&
  {Mikofski}}]{Holmgren2018_pvlib}
{Holmgren}, W.~F., {Hansen}, C.~W., \& {Mikofski}, M.~A. 2018, Journal of Open
  Source Software, 3, 884, \dodoi{https://doi.org/10.21105/joss.00884}

\bibitem[{Hunter(2007)}]{Hunter2007_matplotlib}
Hunter, J.~D. 2007, Computing In Science \& Engineering, 9, 90,
  \dodoi{10.1109/MCSE.2007.55}

\bibitem[{{Jones} {et~al.}(2017){Jones}, {Stenning}, {Ford}, {Wolpert},
  {Loredo}, \& {Dumusque}}]{Jones_2017}
{Jones}, D.~E., {Stenning}, D.~C., {Ford}, E.~B., {et~al.} 2017, arXiv
  e-prints, arXiv:1711.01318.
\newblock \doarXiv{1711.01318}

\bibitem[{{Jurgenson} {et~al.}(2016){Jurgenson}, {Fischer}, {McCracken},
  {Sawyer}, {Szymkowiak}, {Davis}, {Muller}, \& {Santoro}}]{EXPRES_main}
{Jurgenson}, C., {Fischer}, D., {McCracken}, T., {et~al.} 2016, in Society of
  Photo-Optical Instrumentation Engineers (SPIE) Conference Series, Vol. 9908,
  \procspie, 99086T, \dodoi{10.1117/12.2233002}

\bibitem[{{Kanodia} \& {Wright}(2018)}]{Kanodia2018_barycorrpy}
{Kanodia}, S., \& {Wright}, J. 2018, Research Notes of the American
  Astronomical Society, 2, 4, \dodoi{10.3847/2515-5172/aaa4b7}

\bibitem[{{Kanodia} {et~al.}(2018){Kanodia}, {Mahadevan}, {Ramsey},
  {Stefansson}, {Monson}, {Hearty}, {Blakeslee}, {Lubar}, {Bender}, {Ninan},
  {Sterner}, {Roy}, {Halverson}, \& {Robertson}}]{Kanodia2018_HPFfibers}
{Kanodia}, S., {Mahadevan}, S., {Ramsey}, L.~W., {et~al.} 2018, in Society of
  Photo-Optical Instrumentation Engineers (SPIE) Conference Series, Vol. 10702,
  Ground-based and Airborne Instrumentation for Astronomy VII, ed. C.~J.
  {Evans}, L.~{Simard}, \& H.~{Takami}, 107026Q, \dodoi{10.1117/12.2313491}

\bibitem[{{Logsdon} {et~al.}(2018){Logsdon}, {McElwain}, {Gong}, {Liang},
  {Santoro}, {Schwab}, {Bender}, {Blake}, {Halverson}, \&
  {Hearty}}]{NEID_Port_Overview}
{Logsdon}, S.~E., {McElwain}, M.~W., {Gong}, Q., {et~al.} 2018, in Society of
  Photo-Optical Instrumentation Engineers (SPIE) Conference Series, Vol. 10702,
  Ground-based and Airborne Instrumentation for Astronomy VII, 1070267,
  \dodoi{10.1117/12.2312209}

\bibitem[{{Lubin} {et~al.}(2021){Lubin}, {Robertson}, {Stefansson}, {Ninan},
  {Mahadevan}, {Endl}, {Ford}, {Wright}, {Beard}, {Bender}, {Cochran},
  {Diddams}, {Fredrick}, {Halverson}, {Kanodia}, {Metcalf}, {Ramsey}, {Roy},
  {Schwab}, \& {Terrien}}]{Lubin2021}
{Lubin}, J., {Robertson}, P., {Stefansson}, G., {et~al.} 2021, \aj, 162, 61,
  \dodoi{10.3847/1538-3881/ac0057}

\bibitem[{{Maldonado} {et~al.}(2019){Maldonado}, {Phillips}, {Dumusque},
  {Collier Cameron}, {Haywood}, {Lanza}, {Micela}, {Mortier}, {Saar},
  {Sozzetti}, {Rice}, {Milbourne}, {Cecconi}, {Cegla}, {Cosentino}, {Costes},
  {Ghedina}, {Gonzalez}, {Guerra}, {Hern{\'a}ndez}, {Li}, {Lodi}, {Malavolta},
  {Molinari}, {Pepe}, {Piotto}, {Poretti}, {Sasselov}, {San Juan}, {Thompson},
  {Udry}, \& {Watson}}]{Maldonado_2019}
{Maldonado}, J., {Phillips}, D.~F., {Dumusque}, X., {et~al.} 2019, \aap, 627,
  A118, \dodoi{10.1051/0004-6361/201935233}

\bibitem[{{Marchwinski} {et~al.}(2015){Marchwinski}, {Mahadevan}, {Robertson},
  {Ramsey}, \& {Harder}}]{Marchwinski_2015}
{Marchwinski}, R.~C., {Mahadevan}, S., {Robertson}, P., {Ramsey}, L., \&
  {Harder}, J. 2015, \apj, 798, 63, \dodoi{10.1088/0004-637X/798/1/63}

\bibitem[{{Mayor} \& {Queloz}(1995)}]{Mayor_1995}
{Mayor}, M., \& {Queloz}, D. 1995, \nat, 378, 355, \dodoi{10.1038/378355a0}

\bibitem[{McKinney(2010)}]{McKinney2010_pandas}
McKinney, W. 2010, in Proceedings of the 9th Python in Science Conference, ed.
  S.~van~der Walt \& J.~Millman, 51 -- 56

\bibitem[{{Meunier} \& {Lagrange}(2019)}]{Meunier_2019}
{Meunier}, N., \& {Lagrange}, A.~M. 2019, \aap, 625, L6,
  \dodoi{10.1051/0004-6361/201935099}

\bibitem[{{Meunier} {et~al.}(2017){Meunier}, {Lagrange}, \&
  {Borgniet}}]{Meunier_2017b}
{Meunier}, N., {Lagrange}, A.~M., \& {Borgniet}, S. 2017, \aap, 607, A6,
  \dodoi{10.1051/0004-6361/201630328}

\bibitem[{{Miklos} {et~al.}(2020){Miklos}, {Milbourne}, {Haywood}, {Phillips},
  {Saar}, {Meunier}, {Cegla}, {Dumusque}, {Langellier}, {Maldonado},
  {Malavolta}, {Mortier}, {Thompson}, {Watson}, {Cecconi}, {Cosentino},
  {Ghedina}, {Li}, {L{\'o}pez-Morales}, {Molinari}, {Poretti}, {Sasselov},
  {Sozzetti}, \& {Walsworth}}]{Miklos_2020}
{Miklos}, M., {Milbourne}, T.~W., {Haywood}, R.~D., {et~al.} 2020, \apj, 888,
  117, \dodoi{10.3847/1538-4357/ab59d5}

\bibitem[{{Milbourne} {et~al.}(2019){Milbourne}, {Haywood}, {Phillips}, {Saar},
  {Cegla}, {Cameron}, {Costes}, {Dumusque}, {Langellier}, {Latham},
  {Maldonado}, {Malavolta}, {Mortier}, {Palumbo}, {Thompson}, {Watson},
  {Bouchy}, {Buchschacher}, {Cecconi}, {Charbonneau}, {Cosentino}, {Ghedina},
  {Glenday}, {Gonzalez}, {Li}, {Lodi}, {L{\'o}pez-Morales}, {Lovis}, {Mayor},
  {Micela}, {Molinari}, {Pepe}, {Piotto}, {Rice}, {Sasselov}, {S{\'e}gransan},
  {Sozzetti}, {Szentgyorgyi}, {Udry}, \& {Walsworth}}]{Milbourne_2019}
{Milbourne}, T.~W., {Haywood}, R.~D., {Phillips}, D.~F., {et~al.} 2019, \apj,
  874, 107, \dodoi{10.3847/1538-4357/ab064a}

\bibitem[{{National Academies of Sciences, Engineering, and
  Medicine}(2021)}]{astro2020decadal}
{National Academies of Sciences, Engineering, and Medicine}. 2021, Pathways to
  Discovery in Astronomy and Astrophysics for the 2020s (Washington, DC: The
  National Academies Press), \dodoi{10.17226/26141}

\bibitem[{{Pepe} {et~al.}(2014){Pepe}, {Molaro}, {Cristiani}, {Rebolo},
  {Santos}, {Dekker}, {M{\'e}gevand}, {Zerbi}, {Cabral}, {Di Marcantonio},
  {Abreu}, {Affolter}, {Aliverti}, {Allende Prieto}, {Amate}, {Avila},
  {Baldini}, {Bristow}, {Broeg}, {Cirami}, {Coelho}, {Conconi}, {Coretti},
  {Cupani}, {D'Odorico}, {De Caprio}, {Delabre}, {Dorn}, {Figueira}, {Fragoso},
  {Galeotta}, {Genolet}, {Gomes}, {Gonz{\'a}lez Hern{\'a}ndez}, {Hughes},
  {Iwert}, {Kerber}, {Landoni}, {Lizon}, {Lovis}, {Maire}, {Mannetta},
  {Martins}, {Monteiro}, {Oliveira}, {Poretti}, {Rasilla}, {Riva}, {Santana
  Tschudi}, {Santos}, {Sosnowska}, {Sousa}, {Span{\'o}}, {Tenegi}, {Toso},
  {Vanzella}, {Viel}, \& {Zapatero Osorio}}]{ESPRESSO_main}
{Pepe}, F., {Molaro}, P., {Cristiani}, S., {et~al.} 2014, Astronomische
  Nachrichten, 335, 8, \dodoi{12.1002/asna.201312004}

\bibitem[{{Phillips} {et~al.}(2016){Phillips}, {Glenday}, {Dumusque},
  {Buchschacher}, {Cameron}, {Cecconi}, {Charbonneau}, {Cosentino}, {Ghedina},
  {Haywood}, {Latham}, {Li}, {Lodi}, {Lovis}, {Molinari}, {Pepe}, {Sasselov},
  {Szentgyorgyi}, {Udry}, \& {Walsworth}}]{Phillips_2016}
{Phillips}, D.~F., {Glenday}, A.~G., {Dumusque}, X., {et~al.} 2016, in Society
  of Photo-Optical Instrumentation Engineers (SPIE) Conference Series, Vol.
  9912, \procspie, 99126Z, \dodoi{10.1117/12.2232452}

\bibitem[{{Rajpaul} {et~al.}(2015){Rajpaul}, {Aigrain}, {Osborne}, {Reece}, \&
  {Roberts}}]{Rajpaul_2015}
{Rajpaul}, V., {Aigrain}, S., {Osborne}, M.~A., {Reece}, S., \& {Roberts}, S.
  2015, \mnras, 452, 2269, \dodoi{10.1093/mnras/stv1428}

\bibitem[{{Rajpaul} {et~al.}(2020){Rajpaul}, {Aigrain}, \&
  {Buchhave}}]{Rajpaul_2020}
{Rajpaul}, V.~M., {Aigrain}, S., \& {Buchhave}, L.~A. 2020, \mnras, 492, 3960,
  \dodoi{10.1093/mnras/stz3599}

\bibitem[{{Robertson} {et~al.}(2016{\natexlab{a}}){Robertson}, {Bender},
  {Mahadevan}, {Roy}, \& {Ramsey}}]{Robertson_2016}
{Robertson}, P., {Bender}, C., {Mahadevan}, S., {Roy}, A., \& {Ramsey}, L.~W.
  2016{\natexlab{a}}, \apj, 832, 112, \dodoi{10.3847/0004-637X/832/2/112}

\bibitem[{{Robertson} {et~al.}(2014){Robertson}, {Mahadevan}, {Endl}, \&
  {Roy}}]{Robertson_2014}
{Robertson}, P., {Mahadevan}, S., {Endl}, M., \& {Roy}, A. 2014, Science, 345,
  440, \dodoi{10.1126/science.1253253}

\bibitem[{{Robertson} {et~al.}(2019){Robertson}, {Anderson}, {Stefansson},
  {Hearty}, {Monson}, {Mahadevan}, {Blakeslee}, {Bender}, {Ninan}, {Conran},
  {Levi}, {Lubar}, {Cole}, {Dykhouse}, {Kanodia}, {Nitroy}, {Smolsky},
  {Tuggle}, {Blank}, {Nelson}, {Blake}, {Halverson}, {Henderson}, {Kaplan},
  {Li}, {Logsdon}, {McElwain}, {Rajagopal}, {Ramsey}, {Roy}, {Schwab},
  {Terrien}, \& {Wright}}]{Robertson_environment}
{Robertson}, P., {Anderson}, T., {Stefansson}, G., {et~al.} 2019, Journal of
  Astronomical Telescopes, Instruments, and Systems, 5, 015003,
  \dodoi{10.1117/1.JATIS.5.1.015003}

\bibitem[{{Robertson} {et~al.}(2016{\natexlab{b}}){Robertson}, {Hearty},
  {Anderson}, {Stef{\'a}nsson}, {Levi}, {Bender}, {Mahadevan}, {Halverson},
  {Monson}, {Ramsey}, {Roy}, {Schwab}, {Terrien}, {Nelson}, \&
  {Blank}}]{Robertson_thermal}
{Robertson}, P.~M., {Hearty}, F.~R., {Anderson}, T.~B., {et~al.}
  2016{\natexlab{b}}, in \procspie, Vol. 9908, Ground-based and Airborne
  Instrumentation for Astronomy VI, 990862, \dodoi{10.1117/12.2231311}

\bibitem[{{Santos} {et~al.}(2014){Santos}, {Mortier}, {Faria}, {Dumusque},
  {Adibekyan}, {Delgado-Mena}, {Figueira}, {Benamati}, {Boisse}, {Cunha},
  {Gomes da Silva}, {Lo Curto}, {Lovis}, {Martins}, {Mayor}, {Melo}, {Oshagh},
  {Pepe}, {Queloz}, {Santerne}, {S{\'e}gransan}, {Sozzetti}, {Sousa}, \&
  {Udry}}]{Santos_2014}
{Santos}, N.~C., {Mortier}, A., {Faria}, J.~P., {et~al.} 2014, \aap, 566, A35,
  \dodoi{10.1051/0004-6361/201423808}

\bibitem[{{Schwab} {et~al.}(2016){Schwab}, {Rakich}, {Gong}, {Mahadevan},
  {Halverson}, {Roy}, {Terrien}, {Robertson}, {Hearty}, {Levi}, {Monson},
  {Wright}, {McElwain}, {Bender}, {Blake}, {St{\"u}rmer}, {Gurevich},
  {Chakraborty}, \& {Ramsey}}]{NEID_optical}
{Schwab}, C., {Rakich}, A., {Gong}, Q., {et~al.} 2016, in \procspie, Vol. 9908,
  Ground-based and Airborne Instrumentation for Astronomy VI, 99087H,
  \dodoi{10.1117/12.2234411}

\bibitem[{{Schwab} {et~al.}(2018){Schwab}, {Liang}, {Gong}, {Bender}, {Blake},
  {Halverson}, {Harbeck}, {Hearty}, {Hunting}, \& {Jaehnig}}]{NEID_Port_Design}
{Schwab}, C., {Liang}, M., {Gong}, Q., {et~al.} 2018, in Society of
  Photo-Optical Instrumentation Engineers (SPIE) Conference Series, Vol. 10702,
  Ground-based and Airborne Instrumentation for Astronomy VII, 1070271,
  \dodoi{10.1117/12.2314420}

\bibitem[{{Seifahrt} {et~al.}(2016){Seifahrt}, {Bean}, {St{\"u}rmer}, {Gers},
  {Grobler}, {Reed}, \& {Jones}}]{maroonx_main}
{Seifahrt}, A., {Bean}, J.~L., {St{\"u}rmer}, J., {et~al.} 2016, in Society of
  Photo-Optical Instrumentation Engineers (SPIE) Conference Series, Vol. 9908,
  Ground-based and Airborne Instrumentation for Astronomy VI, ed. C.~J.
  {Evans}, L.~{Simard}, \& H.~{Takami}, 990818, \dodoi{10.1117/12.2232069}

\bibitem[{{Stefansson} {et~al.}(2016){Stefansson}, {Hearty}, {Robertson},
  {Mahadevan}, {Anderson}, {Levi}, {Bender}, {Nelson}, {Monson}, {Blank},
  {Halverson}, {Henderson}, {Ramsey}, {Roy}, {Schwab}, \&
  {Terrien}}]{Stefansson_thermal}
{Stefansson}, G., {Hearty}, F., {Robertson}, P., {et~al.} 2016, \apj, 833, 175,
  \dodoi{10.3847/1538-4357/833/2/175}

\bibitem[{{Szentgyorgyi} {et~al.}(2016){Szentgyorgyi}, {Baldwin}, {Barnes},
  {Bean}, {Ben-Ami}, {Brennan}, {Budynkiewicz}, {Chun}, {Conroy}, {Crane},
  {Epps}, {Evans}, {Evans}, {Foster}, {Frebel}, {Gauron}, {Guzm{\'a}n}, {Hare},
  {Jang}, {Jang}, {Jordan}, {Kim}, {Kim}, {Mendes de Oliveira},
  {Lopez-Morales}, {McCracken}, {McMuldroch}, {Miller}, {Mueller}, {Oh},
  {Onyuksel}, {Ordway}, {Park}, {Park}, {Park}, {Paxson}, {Phillips},
  {Plummer}, {Podgorski}, {Seifahrt}, {Stark}, {Steiner}, {Uomoto},
  {Walsworth}, \& {Yu}}]{GCLEF_main}
{Szentgyorgyi}, A., {Baldwin}, D., {Barnes}, S., {et~al.} 2016, in Society of
  Photo-Optical Instrumentation Engineers (SPIE) Conference Series, Vol. 9908,
  \procspie, 990822, \dodoi{10.1117/12.2233506}

\bibitem[{{Thompson} {et~al.}(2020){Thompson}, {Watson}, {Haywood}, {Costes},
  {de Mooij}, {Collier Cameron}, {Dumusque}, {Phillips}, {Saar}, {Mortier},
  {Milbourne}, {Aigrain}, {Cegla}, {Charbonneau}, {Cosentino}, {Ghedina},
  {Latham}, {L{\'o}pez-Morales}, {Micela}, {Molinari}, {Poretti}, {Sozzetti},
  {Thompson}, \& {Walsworth}}]{Thompson_2020}
{Thompson}, A.~P.~G., {Watson}, C.~A., {Haywood}, R.~D., {et~al.} 2020, arXiv
  e-prints, arXiv:2004.09830.
\newblock \doarXiv{2004.09830}

\bibitem[{{Thompson} {et~al.}(2016){Thompson}, {Queloz}, {Baraffe}, {Brake},
  {Dolgopolov}, {Fisher}, {Fleury}, {Geelhoed}, {Hall}, {Gonz{\'a}lez
  Hern{\'a}ndez}, {ter Horst}, {Kragt}, {Navarro}, {Naylor}, {Pepe},
  {Piskunov}, {Rebolo}, {Sand er}, {S{\'e}gransan}, {Seneta}, {Sing},
  {Snellen}, {Snik}, {Spronck}, {Stempels}, {Sun}, {Santana Tschudi}, \&
  {Young}}]{HARPS3_main}
{Thompson}, S.~J., {Queloz}, D., {Baraffe}, I., {et~al.} 2016, in Society of
  Photo-Optical Instrumentation Engineers (SPIE) Conference Series, Vol. 9908,
  \procspie, 99086F, \dodoi{10.1117/12.2232111}

\bibitem[{{van der Walt} {et~al.}(2011){van der Walt}, {Colbert}, \&
  {Varoquaux}}]{vanderWalt2011_numpy}
{van der Walt}, S., {Colbert}, S.~C., \& {Varoquaux}, G. 2011, Computing in
  Science and Engineering, 13, 22, \dodoi{10.1109/MCSE.2011.37}

\bibitem[{{Virtanen} {et~al.}(2020){Virtanen}, {Gommers}, {Oliphant},
  {Haberland}, {Reddy}, {Cournapeau}, {Burovski}, {Peterson}, {Weckesser},
  {Bright}, {van der Walt}, {Brett}, {Wilson}, {Millman}, {Mayorov}, {Nelson},
  {Jones}, {Kern}, {Larson}, {Carey}, {Polat}, {Feng}, {Moore}, {Vand erPlas},
  {Laxalde}, {Perktold}, {Cimrman}, {Henriksen}, {Quintero}, {Harris},
  {Archibald}, {Ribeiro}, {Pedregosa}, {van Mulbregt}, \& {SciPy 1. 0
  Contributors}}]{Virtanen2020_scipy}
{Virtanen}, P., {Gommers}, R., {Oliphant}, T.~E., {et~al.} 2020, Nature
  Methods, 17, 261, \dodoi{10.1038/s41592-019-0686-2}

\bibitem[{{Wolszczan}(1994)}]{Wolszczan_1994}
{Wolszczan}, A. 1994, Science, 264, 538, \dodoi{10.1126/science.264.5158.538}

\bibitem[{{Wolszczan} \& {Frail}(1992)}]{Wolszczan_1992}
{Wolszczan}, A., \& {Frail}, D.~A. 1992, \nat, 355, 145,
  \dodoi{10.1038/355145a0}

\bibitem[{{Wright} \& {Eastman}(2014)}]{WrightEastman2014_barycorr}
{Wright}, J.~T., \& {Eastman}, J.~D. 2014, \pasp, 126, 838,
  \dodoi{10.1086/678541}

\bibitem[{{Wright} \& {Kanodia}(2020)}]{WrightKanodia2020}
{Wright}, J.~T., \& {Kanodia}, S. 2020, \psj, 1, 38, \dodoi{10.3847/PSJ/ababa4}

\end{thebibliography}
\bibliographystyle{aasjournal}

\end{document}